\documentclass[10pt]{article}
\usepackage{authblk}
\usepackage{etoc}
\usepackage{multirow}
\usepackage{geometry}
\usepackage{float}
\usepackage{amsmath} 
\usepackage[pdfencoding=auto, psdextra]{hyperref}
\usepackage{nameref} 
\usepackage[dvipdfmx]{graphicx}
\usepackage[]{algorithm2e}
\usepackage{url} 
\usepackage{subcaption}
\usepackage[T1]{fontenc}
\usepackage{array}
\usepackage{makecell}
\usepackage{color}
\usepackage{amsmath}
\usepackage[normalem]{ulem}
\usepackage{pbox,tikz,calc,ragged2e}
\usepackage{csquotes}
\usepackage{soul,xcolor}
\usepackage[shortlabels]{enumitem}
\usepackage{amsfonts}

\usepackage{tcolorbox}
\usepackage{comment}
\usepackage{bm}

\usepackage{etoolbox}
\usepackage{environ}

\newtoggle{bibdoi}
\newtoggle{biburl}
\makeatletter

\DeclareMathOperator*{\E}{\mathbb{E}}

\usepackage[symbol]{footmisc}

\providecommand{\keywords}[1]{\textbf{\textit{Keywords:}} #1}

\usepackage{titlesec} 
\usepackage{apacite}
\titleformat{\section}
  {\normalfont\fontfamily{phv}\fontsize{10}{19}\bfseries}{\thesection}{1em}{}
\titleformat{\subsection}
  {\normalfont \fontsize{10}{17}\itshape}{\thesubsection}{1em}{}
  
\undef{\APACrefURL}
\undef{\endAPACrefURL}
\undef{\APACrefDOI}
\undef{\endAPACrefDOI}

\long\def\collect@url#1{\global\def\bib@url{#1}}
\long\def\collect@doi#1{\global\def\bib@doi{#1}}
\newenvironment{APACrefURL}{\global\toggletrue{biburl}\Collect@Body\collect@url}{\unskip\unskip}
\newenvironment{APACrefDOI}{\global\toggletrue{bibdoi}\Collect@Body\collect@doi}{}

\AtBeginEnvironment{thebibliography}{
 \pretocmd{\PrintBackRefs}{%
  \iftoggle{bibdoi}
    {\iftoggle{biburl}{\unskip\unskip doi:\bib@doi}{}}
    {\iftoggle{biburl}{\bib@url}{}}
  \togglefalse{bibdoi}\togglefalse{biburl}%
  }{}{}
}
  
\begin{document}
\title{Joint Estimation of the Non-parametric Transitivity and Preferential Attachment Functions in Scientific Co-authorship Networks}
\author[1,2]{Masaaki Inoue\thanks{Corresponding author: minoue@sys.i.kyoto-u.ac.jp}\footnote{These authors contributed equally.}}
\newcommand\CoAuthorMark{\footnotemark[\arabic{footnote}]}
\author[2]{Thong Pham\protect\CoAuthorMark}
\author[1,2]{Hidetoshi Shimodaira}
\affil[1]{Kyoto University}
\affil[2]{RIKEN Center for AIP}

\date{}
\maketitle


\begin{abstract}
We propose a statistical method to estimate simultaneously the non-parametric transitivity and preferential attachment functions in a growing network, in contrast to conventional methods that either estimate each function in isolation or assume some functional form for them. Our  model is shown to be a good fit to two real-world co-authorship networks and be able to bring to light intriguing details of the preferential attachment and transitivity phenomena that would be unavailable under traditional methods. We also introduce a method to quantify the amount of contributions of those phenomena in the growth process of a network based on the probabilistic dynamic process induced by the model formula. Applying this method, we found that transitivity dominated preferential attachment in both co-authorship networks. This suggests the importance of indirect relations in scientific creative processes. The proposed methods are implemented in the R package \texttt{FoFaF}.
\end{abstract}

\section{Introduction}\label{sec:intro}

Science has never been more collaborative. In this era that has been witnessing an unprecedented explosion of multi-author scholarly articles~\cite{citation_1}, collaboration has become more and more important in the path to scientific success~\cite{collaboration_science_1,citation_2}. Promising ideas from numerous analytic fields, including complex network theory, statistics, and informetrics, have been weaved together to understand this collaborative nature of science~\cite{scisci2,scisci1}.

One of the early attempts to analyze the formative process of scientific collaborations came from physics when Newman proposed a non-parametric method to estimate the preferential attachment (PA) and transitivity functions from scientific collaboration networks~\cite{newman2001clustering}. PA~\cite{price2,matthew_effect,price,barabasi-albert} and transitivity~\cite{fritz,holland_1970,holland_1971,holland_1975} are two of the most fundamental mechanisms of network growth. On the one hand, PA is a phenomenon concerning  the first-order structure of a network. In PA, the higher the number of co-authors a scientist already has, the more collaborators they will form. On the other hand, transitivity concerns the second-order structure: co-authors of co-authors are likely to collaborate. Newman's method is non-parametric in the sense that it does not assume any forms for either the PA or transitivity function. The method, however, considers each phenomenon in isolation and thus completely ignores any entanglements of the two phenomena, which are entirely plausible in real-world networks.

Apart from this non-parametric-in-isolation approach, a joint-estimation approach, in which the two phenomena are considered simultaneously, has been attempted recently~\cite{rsiena_slovenian_2012,rsiena_transitivity_1,rsiena_transitivity_2}, all under the framework of stochastic actor-based models~\cite{stochastic_actor_1}. This approach is, however, inherently parametric: it assumes the forms of the PA and transitivity functions a \emph{priori}, thus risks losing fine details of the two phenomena, details that are difficult to be captured by any parametric functional forms.

We argue that the ideal method, whenever possible, should combine the best of both worlds: it should consider both phenomena simultaneously, and it should not assume any functional forms for them.

Our main contributions are three-fold. In our first contribution, we propose a network growth model that combines non-parametric PA and transitivity functions and derive an efficient Minorize-Maximization (MM) algorithm~\cite{MM} to estimate them simultaneously. This iterative algorithm is guaranteed to increase the model's log-likelihood per iteration. We demonstrate through simulated examples that our approach are capable of capturing complex details of PA and transitivity, while the conventional approaches cannot (cf. Fig.~\ref{fig:simulation}). We also perform a systematic simulation to confirm the performance of our algorithm.

In our second contribution, we suggest a method to quantify the amount of contributions of PA and transitivity in the growth process of a network. Our quantification exploits the probabilistic dynamic process induced by the network growth formula and can be easily extended to other network growth mechanisms. 

In our third contribution, we apply the proposed methods to two real-world co-authorship networks and uncover some interesting properties that would be unavailable under conventional approaches. In particular, as contrast to the typical power-law functional form assumption, the transitivity effect seems to be highly non-power-law. We also found that transitivity dominated PA in the growth processes of both networks. This suggests the importance of indirect relations in scientific creative processes: it does matter who your collaborators collaborate with. All the proposed methods are implemented in the R package \texttt{FoFaF}.

The rest of the paper is organized as follows. The proposed method is discussed in details in Section~\ref{sec:method}. In Section~\ref{sec:quantify}, we discuss how to exploit the probabilistic dynamic process imposed by the model formula to sensibly quantify the amount of contributions of PA and transitivity. We apply the proposed method to two real-world collaboration networks and discuss the results in Section~\ref{sec:result}. Concluding remarks are given in Section~\ref{sec:conclusion}.

\section{Proposed Method}\label{sec:method}
We first review briefly the history of PA and transitivity modelling and then describe our network growth model that incorporates non-parametric PA and transitivity functions. We also explain its relation to some conventional network models. We then discuss maximum partial likelihood estimation for the model and provide two simulated examples to demonstrate how our method works. We conclude the section with a systematic simulation to investigate the performance of the proposed method.
\subsection{PA and transitivity modelling}\label{sec:sub_method_in_the_literature}
The notion of a rich-get-richer phenomenon has its root in the theoretical works of Yule~\cite{yule} and Simon~\cite{simon}. Its status as a fundamental process in informetrics was cemented by revolutionary works of Merton~\cite{matthew_effect} and Price~\cite{price2,price}. The term ``preferential attachment'' was coined by Barab\'{a}si and Albert when they re-discovered the mechanism in the context of complex networks~\cite{barabasi-albert}. 

In PA, the probability a node with degree $k$ receives a new edge is proportional to its PA function $A_k$. When $A_k$ is an increasing function on average, the PA effect exists: a node with a high degree $k$ is more likely to receive more new connections. To estimate the PA phenomenon in a network is to estimate the function $A_k$ given that network's growth data. Various non-parametric approaches~\cite{newman2001clustering,pham2} and parametric ones~\cite{massen,Gomez} have been proposed. In parametric methods, power-law function forms, e.g., $A_k = (k + 1)^\alpha$, are often employed~\cite{krapi}.

Transitivity started out as a concept in psychology~\cite{fritz} and was developed theoretically in the framework of social network analysis by Holland and Leinhardt in the 1970s~\cite{holland_1970,holland_1971,holland_1975}. It was introduced to the informetrician's modelling toolbox in 2001 when Newman provided a heuristic method to estimate the transitivity function in real-world co-authorship networks~\cite{newman2001clustering} and Snijders introduced his now-famous stochastic actor-based models that include transitivity as a network formation mechanism~\cite{stochastic_actor_1}.

In transitivity, the probability that a pair of two nodes with $b$ common neighbors is proportional to the transitivity function $B_{b}$. When $B_b$ is an increasing function on average, the transitivity effect is at play: the more common neighbors a pair of nodes share, the easier for them to connect. Similar to the case of PA, non-parametric approaches~\cite{newman2001clustering} and parametric approaches~\cite{rsiena_slovenian_2012,rsiena_transitivity_1,rsiena_transitivity_2} have been proposed to estimate $B_b$ from observed network data.

We re-emphasize that all existing methods either consider PA or transitivity in isolation, or are of a parametric nature.  

\subsection{Proposed network model}\label{sec:sub_method_model}
 Our model can be viewed as a discrete Markov model, which is a popular framework in social network modeling~\cite{holland_1977}. Let $G_{t}$ denote the network at time $t$. Starting from a seed network $G_{0}$, at each time-step $t = 1,\cdots, T$, $v(t)$ new nodes and $m(t)$ new edges are added to $G_{t-1}$ to form $G_{t}$. In particular, at the onset of time-step $t$, let $k_i(t)$ denote the degree of node $i$ and $b_{ij}(t)$ the number of common neighbors between nodes $i$ and $j$ in $G_{t-1}$. Our model dictates that the probability that a new edge emerges between node $i$ and node $j$ at time step $t$ is independent of other new edges at that time and is equal to
\begin{equation}
P_{ij}(t) \propto A_{k_i(t)}A_{k_j(t)}B_{b_{ij}(t)},\label{eq:probability}
\end{equation}
where $A_k$ is the PA function of the degree $k$ and $B_b$ is the transitivity function of the number of common neighbors $b$. In other words, the un-ordered pair of the two ends $(i,j)$ of a new edge follows a categorical distribution over all un-ordered pairs of nodes existing at time $t$. Each pair's weight is proportional to the product of PA and transitivity values of that pair at $t$. Thus this formulation can capture simultaneously PA and transitivity effects.

Suppose that the joint distribution of $v(t)$, $m(t)$, and $G_{0}$ is governed by some parameter vector $\boldsymbol{\theta}$. We make a standard assumption, which is virtually employed in all network models, that $\boldsymbol{\theta}$ is independent of $A_{k}$ and $B_{b}$. As we shall see later, this independence assumption enables a partial likelihood approach in which one can ignore $\boldsymbol{\theta}$ in estimating $A_k$ and $B_b$. Next we discuss the relation between the model in Eq.~(\ref{eq:probability}) and models in the literature.

\subsection{Related models}\label{sec:sub_method_rela}
As explained earlier, while there are models that either include a non-parametric $A_k$ function~\cite{pham2} or a non-parametric $B_b$ function~\cite{newman2001clustering}, Eq.~(\ref{eq:probability}) is the first to combine both non-parametric functions. It includes as special cases some well-known complex network models, such as the Barab\'{a}si-Albert model~\cite{barabasi-albert} or the Erd\"{o}s-R\'{e}nyi-with-growth model~\cite{grow-random}. 

The well-known stochastic actor-based model~\cite{stochastic_actor_1,stochastic_actor_2,rsiena} has been employed in studies of scientific co-authorship networks~\cite{rsiena_slovenian_2012,rsiena_transitivity_1,rsiena_transitivity_2}. It is, however, not clear how to convert the PA and transitivity functions in our probabilistic setting to those in the setting of stochastic actor-based model, since the two models are defined differently. We note that the PA and transitivity phenomena are virtually modelled in a parametric manner in the stochastic actor-based model.

One key assumption of the model in Eq.~(\ref{eq:probability}) is that $A_k$ and $B_b$ do not depend on $t$, i.e., they stay unchanged throughout the growth process. While this time invariability assumption is standard and employed in all the network models mentioned previously, there is a growing body of models departing from it. A time-varying $A_k$ has been discussed in the context of citation networks~\cite{gabor,wang_measuring,temporal}, while different parametric forms for such $A_k$ are studied by Medo~\cite{medo_time_varying}. More recently, the \textrm{R} package \texttt{tergm}~\cite{tergm} allows the estimation of time-varying parametric PA and transitivity functions. There is, however, no existing work that employs time-varying and non-parametric modelling simulataneously, presumably for the reason that a huge amount of data is probably needed in such a model. It is likely that in practical situations one always has to choose between non-parametric modelling and time-varying modelling. We demonstrate in Section~\ref{sec:sub_result_diag} that the time invariability assumption do indeed hold in all real-world networks analyzed in this paper.

\subsection{Maximum Partial Likelihood Estimation}\label{sec:sub_method_mle}
Here we describe how to estimate the parameters of the model in Eq.~(\ref{eq:probability}). Denote $D = \{G_{0},G_{1},\cdots, G_{T}\}$ the observed data, and let $\bm{A} = [A_0, A_1,\ldots, A_{k_{max}}]$ with $A_k>0$ be the PA function and $\bm{B} = [B_0, B_1,\ldots, B_{b_{max}}]$ with $B_b>0$ be the transitivity function. Here $k_{max}$ is the maximum degree and $b_{max}$ is the maximum number of common neighbors between a pair of nodes. Given $D$, our goal is to estimate $\bm{A}$ and $\bm{B}$ without assuming any specific functional forms, an approach we call ``non-parametric''.

With the independence assumption stated in the previous section, the part of the log-likelihood that contains $\bm{A}$ and $\bm{B}$ and the part of the log-likelihood that contains $\boldsymbol{\theta}$ are separable, i.e., $L(\bm{A},\bm{B},\bm{\theta}|D) = L(\bm{A},\bm{B}|D)+ L(\bm{\theta}|D)
$ holds, where $L$ denotes the log-likelihood function. This allows us to ignore $\bm{\theta}$ in estimating $A_k$ and $B_b$. Starting from Eq.~(\ref{eq:probability}), with some calculations we arrive at
\begin{align}
L(\bm{A},\bm{B}|D) = &\sum_{t=1}^{T}\sum_{k_1=0 }^{k_{max}}\sum_{k_2=k_1}^{k_{max}}\sum_{b=0}^{b_{max}} m_{k_1,k_2,b}(t) \log{A_{k_1}A_{k_2}B_{b}} - \nonumber \\
&\sum_{t=1}^{T}m(t) \log \left( \sum_{k_1 = 0}^{k_{max} }\sum_{k_2=k_1}^{k_{max} }\sum_{b=0}^{b_{max}} n_{k_1,k_2,b}(t) {A_{k_1}A_{k_2}B_{b}} \right), \label{eq:likelihood2}
\end{align}
where  
 $n_{k_1,k_2,b}(t)$ is the number of node pairs $(i,j)$ that satisfy $(k_{i}(t), k_{j}(t), b_{ij}(t)) = (k_1, k_2, b)$ with $k_1 \le k_2$ at time-step $t$, and
 $m_{k_1,k_2,b}(t)$ is the number of new edges between such node pairs. The number of new edges at time-step $t$ is then expressed as $m(t) = \sum_{k_1=0 }^{k_{max} }\sum_{k_2=k_1}^{k_{max} }\sum_{b=0}^{b_{max}} m_{k_1,k_2,b}(t)$.

Although analytically maximizing $L(\bm{A},\bm{B}|D)$ is intractable, we can derive an efficient MM algorithm that iteratively updates $\bm{A}$ and $\bm{B}$. See Appendix A for its derivation. We also denote the final result of the algorithm $\hat{\bm{A}}$ and $\hat{\bm{B}}$, estimates of $\bm{A}$ and $\bm{B}$. 

\subsection{Illustrated examples}\label{sec:sub_method_illu}
We demonstrate the effectiveness of our method in two examples. In the first example, we simulate a network using Eq.~(\ref{eq:probability}) with $A_k = 3\log(\max{k,1})^\alpha + 1$ and $B_b = 3\log(\max{b,1})^\alpha + 1$. This functional form, which deviates substantially from the power-law form, has been used to demonstrate the working of non-parametric PA estimation methods~\cite{pham2}. The network has a total of $N = 1000$ nodes. At each time-step, one new node is added to the network with $m(t)= 5$ new edges. In the second example, we first estimate $A_k$ and $B_b$ by applying our proposed method to a real-world co-authorship network between authors in statistics journals (cf.~Section~\ref{sec:result}), and then use these parameter values for simulating a network based on Eq.~(\ref{eq:probability}). In the process, we kept the initial condition and the number of new nodes and new edges at each time-step exactly as what were observed in the real-world network.

We apply three estimation methods to each simulated network. The first is our proposed method, which jointly estimates the non-parametric functions $A_k$ and $B_b$. The second is a joint parametric method, which jointly estimates PA and transitivity using the simplistic functional forms $A_k = (k + 1)^\alpha$ and $B_b = (b+1)^\alpha$. This parametric formation is used widely in various PA and transitivity estimation methods~\cite{massen,Gomez}. The third method ignores the joint existence of PA and transitivity: it consists of two sub-methods: the first one is a non-parametric method for estimating PA in isolation~\cite{pham2}, and the second one is a maximum likelihood version of a non-parametric method for estimating transitivity in isolation~\cite{newman2001clustering}. 

\begin{figure*}[htbp]
    \centering
        \includegraphics[width=\textwidth]{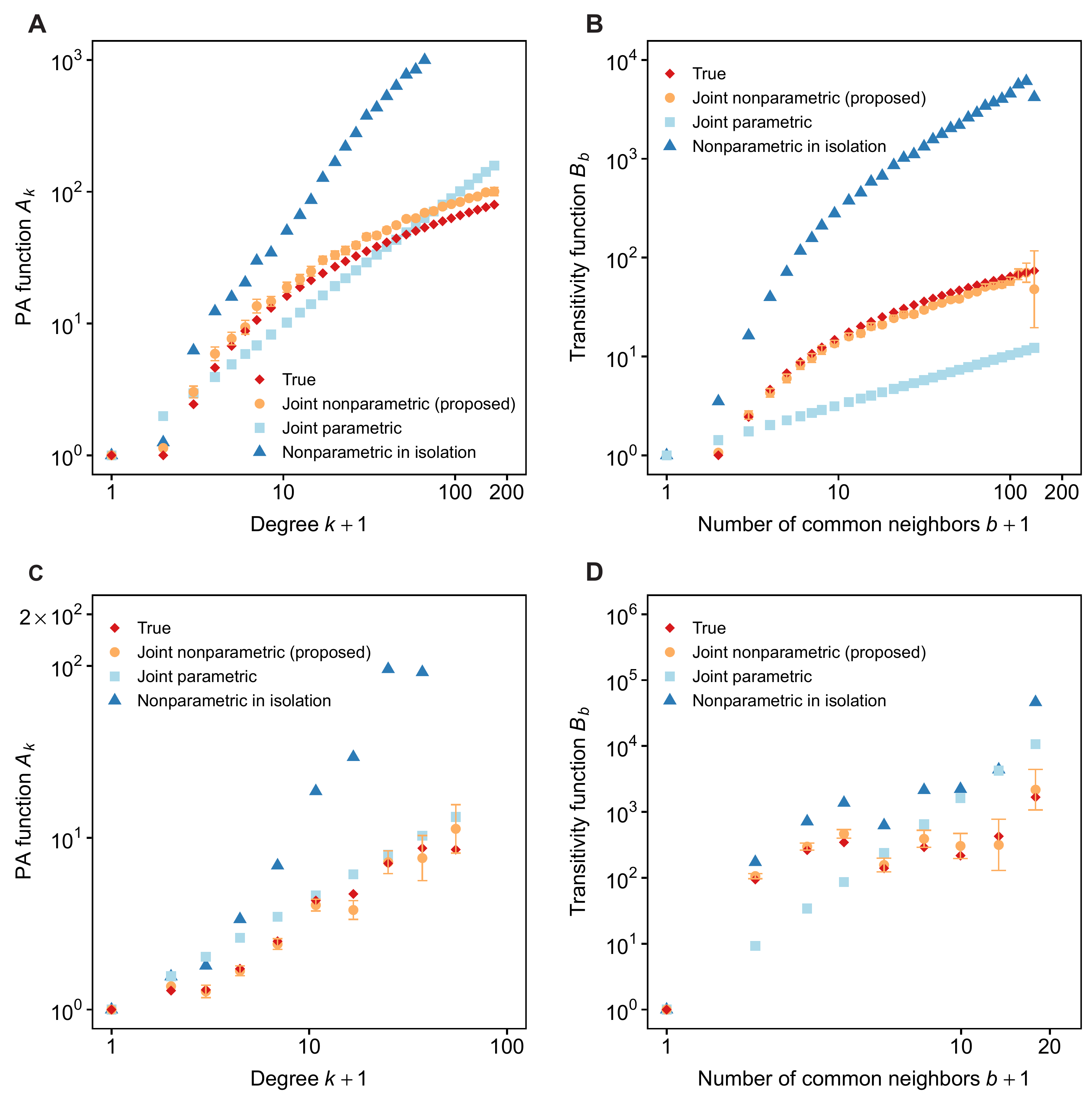}
    \caption{Proposed method compared with conventional methods in estimating the PA and transitivity functions in two simulated networks. \textbf{A}, \textbf{B}: the estimated PA and transitivity functions from a simulated network with $A_k = 3\log(\max{k,1})^\alpha + 1$ and $B_b = 3\log(\max{b,1})^\alpha + 1$. \textbf{C}, \textbf{D}: the estimated PA and transitivity functions from a simulated network in which the true $A_k$ and $B_b$ are the $A_k$ and $B_b$ estimated from a real-world co-authorship network between authors in statistics journals. The interval at each point of the proposed method represents the standard deviation obtained as a by-product of the maximum likelihood estimation. In both networks, the proposed method successfully captured the fine details of PA and transitivity.}\label{fig:simulation}
\end{figure*}

The results are shown in Fig.~\ref{fig:simulation}. In both examples, while the joint parametric method somehow succeeded in obtaining the general trends of $A_k$ and $B_b$, it failed to capture the deviations from the power-law form in the two functions. The non-parametric-in-isolation method grossly over-estimated both PA and transitivity mechanisms, due to its complete disregard of their joint existence. The proposed method worked reasonably well, succeeding in capturing the PA and transitivity functions in fine details.

\subsection{Simulation study}\label{sec:sub_method_sim}
We perform a systematic simulation to evaluate how well the proposed method can estimate $A_k$ and $B_b$. We choose $A_k = (k+1)^\alpha$ and $B_b = (b+1)^\beta$ as the true functions. This power-law functional form has been used in previous simulation studies of PA estimation methods~\cite{pham2,pham_jss}. We consider five values ($0$, $0.5$, $1$, $1.5$, and $2$) for the exponent $\alpha$ and six values ($0$, $0.5$, $1$, $1.5$, $2$, $2.5$, and $3$) for the exponent $\beta$. These are the ranges of $\alpha$ and $\beta$ observed in Section~\ref{sec:sub_result_estimate}. For each combination of $\alpha$ and $\beta$, we simulate $10$ networks. In each network, the total number of nodes is $1000$ and there are five new edges at each time-step. 

For each simulated network, we first estimate $A_k$ and $B_b$ as described in the previous section and then fit $(k+1)^\alpha$ and $(b+1)^\beta$ to the estimation results to find the estimates of $\alpha$ and $\beta$. In other words, we indirectly measure how well $A_k$ and $B_b$ are estimated by looking at how well $\alpha$ and $\beta$ are estimated: if the estimates of $\alpha$ and $\beta$ are good, the estimations of $A_k$ and $B_b$ are likely successful, too. 
 
\begin{figure*}[htbp]
    \centering
        \includegraphics[width=0.75\textwidth]{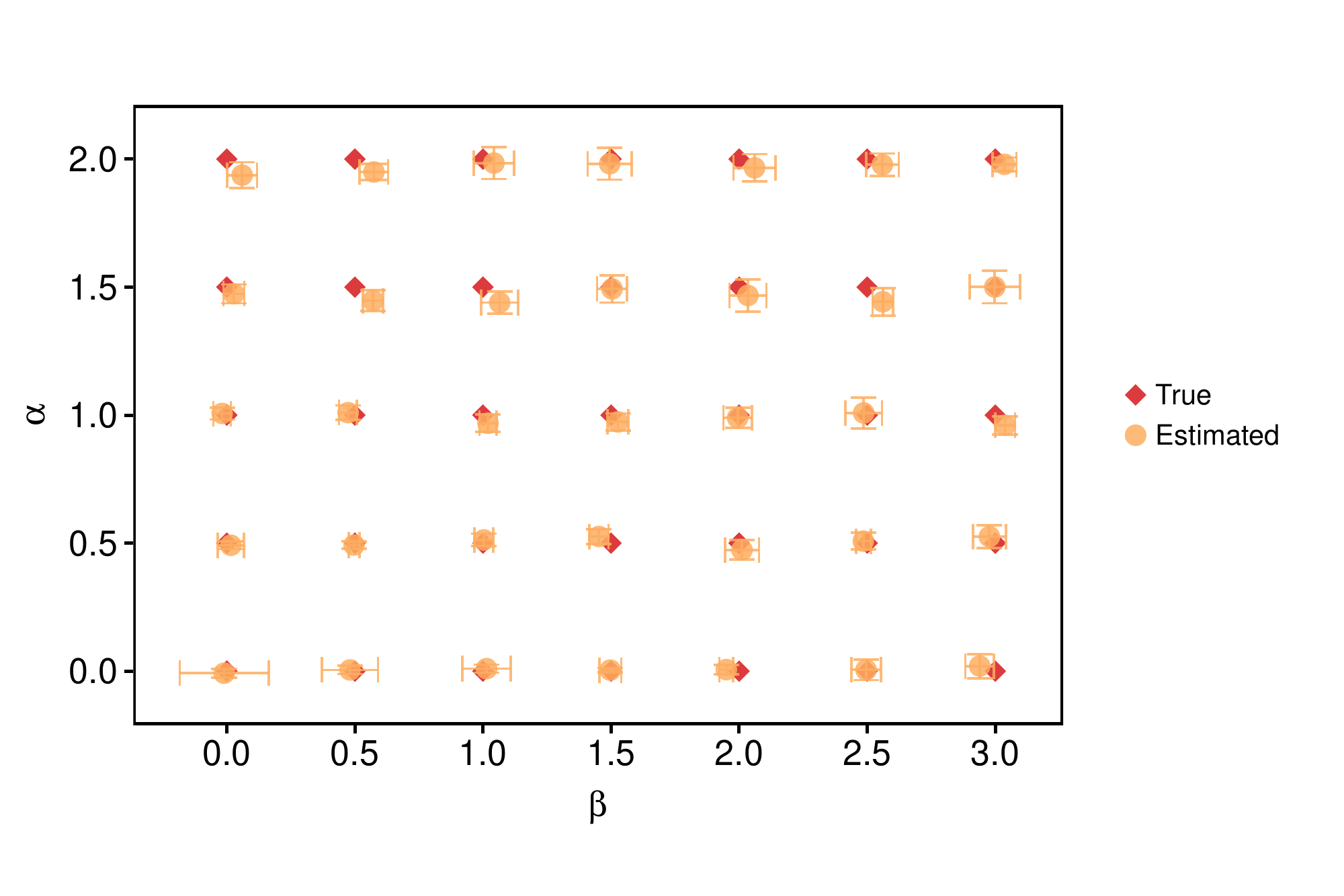}
    \caption{True and estimated exponents $\alpha$ and $\beta$ from the power-law forms $A_k = (k + 1)^\alpha$ and $B_b = (b + 1)^\beta$. The exponents are estimated by a two-step procedure: first $A_k$ and $B_b$ are estimated jointly by the proposed method, then $(k + 1)^\alpha$ and $(b + 1)^\beta$ are fitted to the estimated results by least square. Each estimated point is the mean of the results of $10$ simulations, with the error bars display the standard errors of the mean.} \label{fig:sim_error}
\end{figure*}

Figure~\ref{fig:sim_error} shows the true and estimated values of $\alpha$ and $\beta$. The proposed method successfully recovers $\alpha$ and $\beta$ in all combinations. This implies that the estimation of $A_k$ and $B_b$ went well.

\section{Quantifying the amount of contributions of PA and transitivity}\label{sec:quantify}
Our model leads to a simple answer to a previously-unraised yet fascinating question: how can one compare the amount of contributions of PA and transitivity in the growth process of a network? To the best of our knowledge, no one has attempted to quantify the amount of contributions of different network growth mechanisms. To answer this question, one must find a meaningful way to define the amount of contributions so that they are computable and comparable. We achieve this by considering the dynamic process expressed in Eq.~(\ref{eq:probability}). This probabilistic dynamic process suggests that the variability of the PA/transitivity values in the set of node pairs is a sensible measure for the amount of contribution of PA/transitivity. 

Let us define the amount of contributions of PA and transitivity at time-step $t$. Denote them as $s_{\text{PA}}(t)$ and $s_{\text{trans}}(t)$, respectively. Taking logarithm of both sides of Eq.~(\ref{eq:probability}), one gets:
\begin{equation}
\log_2 P_{ij}(t) = \log_2[A_{k_{i}(t)}A_{k_j(t)}] + \log_2 B_{b_{ij}(t)} - C(t),\label{eq:log_probability}
\end{equation}
with $C(t) = \log_2 \sum_{i,j}A_{k_{i}(t)}A_{k_j(t)}B_{b_{ij}(t)}$ is the logarithm of the normalizing constant at time-step $t$ and is independent of $i$ and $j$. Equation~(\ref{eq:log_probability}) implies that, looking locally at a node pair $(i,j)$, PA and transitivity contribute to $\log_2 P_{ij}(t)$ by the amounts of $\log_2[A_{k_{i}(t)}A_{k_j(t)}]$ and $\log_2 B_{b_{ij}(t)}$, respectively; the amount of contribution is measured by $\log_2$ fold-changes.

What is important globally is, however, the relative sizes of all $\log_2[A_{k_{i}(t)}A_{k_j(t)}]$ and $\log_2 B_{b_{ij}(t)}$ at that time-step $t$. For example, consider the case when $A_k = 1,\forall k$. In this case, the the value of $\log_2[A_{k_{i}(t)}A_{k_j(t)}]$ will be the same for all node pairs, and thus PA would have no role in determining which pair would get a new edge. By considering the case when $B_b = 1,\forall b$, one can see that the same reasoning should also apply to $\log_2 B_{ij}(t)$. 

This observation prompts us to define $s_{\text{PA}}(t)$ and $s_{\text{trans}}(t)$ as the standard deviations of $\log_2[A_{k_{i}(t)}A_{k_j(t)}]$ and $\log_2 B_{b_{ij}(t)}$, respectively, when $(i,j)$ is sampled based on Eq.~(\ref{eq:probability}). Let $U(t)$ be the set formed by all node pairs $(i,j)$ that exist at time-step $t$. The probability $P_{ij}(t)$ in Eq.~(\ref{eq:probability}) can be written explicitly as $$P_{ij}(t) = A_{k_{i}(t)}A_{k_j(t)}B_{b_{ij}(t)} /\sum_{(i,j)\in U(t)} A_{k_{i}(t)}A_{k_j(t)}B_{b_{ij}(t)}.$$
The aforementioned standard deviations can be calculated as follows.
\begin{align}
s_{\text{PA}}(t) &:= \left(\sum_{(i,j)\in U(t)}P_{ij}(t)\left(\log_2[A_{k_{i}(t)}A_{k_j(t)}] - E_{\text{PA}}(t)\right)^2\right)^{1/2},\label{eq:dynamic_PA} \\
s_{\text{trans}}(t) &:= \left(\sum_{(i,j)\in U(t)}P_{ij}(t)\left(\log_2 B_{b_{ij}(t)} - E_{\text{trans}}(t)\right)^2\right)^{1/2},\label{eq:dynamic_trans}
\end{align}
in which $E_{\text{PA}}(t):=\sum_{(i,j) \in U(t)}P_{ij}(t) \log_2[A_{k_{i}(t)}A_{k_j(t)}]$, and $E_{\text{trans}}(t):=\sum_{(i,j) \in U(t)}P_{ij}(t)\log_2 B_{b_{ij}(t)}$. Although $A_k$ and $B_b$ are only defined up to multiplicative constants, the standard deviations of $\log_2[A_{k_{i}(t)}A_{k_j(t)}]$ and $\log_2 B_{b_{ij}(t)}$ are invariant to constant factors in $A_k$ and $B_b$, and thus $s_{\text{PA}}(t)$ and $s_{\text{trans}}(t)$ are well-defined. The use of base-2 logarithms allows us to interpret $s_{\text{PA}}(t)$ and $s_{\text{trans}}(t)$ as $\log_2$ fold-changes; a contribution value of $s$ indicates  a change of the probability $2^s$ times in Eq.~(\ref{eq:probability}). We also note that, although $A_k$ and $B_b$ are assumed to be time-invariant, $k_{i}(t)$, $b_{ij}(t)$, and $P_{ij}(t)$ change over time, thus leading to dynamic nature of $s_{\text{PA}}(t)$ and $s_{\text{trans}}(t)$.
 
In real-world situations, what are available to us is not the true values $\bm{A}$ and $\bm{B}$, but only their estimates $\hat{\bm{A}}$ and $\hat{\bm{B}}$. We plug these estimates into Eqs.~(\ref{eq:dynamic_PA}) and~(\ref{eq:dynamic_trans}) to obtain $\hat{s}_{\text{PA}}(t)$ and $\hat{s}_{\text{trans}}(t)$, estimates of $s_{\text{PA}}(t)$ and $s_{\text{trans}}(t)$.


The requirement that $(i,j)$ is sampled from Eq.~(1) is needed to faithfully reflect the probabilistic dynamic process and leads to the following interpretation of $s_{\text{PA}}(t)$ and $s_{\text{trans}}(t)$. Assume that at some time-step $t$ we observed $m(t) \ge 2$ new edges whose end points are $(i_{1},j_{1}),\cdots,(i_{m(t)},j_{m(t)})$. Consider the sample standard deviation of $\log_{2}(B_{b_{i_{l}j_{l}}(t)})$ for $l = 1,\cdots,m(t)$, which is defined as 
$$h_{\text{trans}}(t) := \left(\frac{1}{m(t) - 1}\sum_{l=1}^{m(t)}\log_{2}(B_{b_{i_{l}j_{l}}(t)})^2 - \frac{1}{m(t)(m(t)-1)}\left(\sum_{l=1}^{m(t)}\log_{2}(B_{b_{i_{l}j_{l}}(t)})\right)^2\right)^{1/2}.$$
Similarly, define $h_{\text{PA}}(t)$ as the sample standard deviation of $\log_{2}(A_{k_{i_{l}}(t)}A_{k_{j_{l}}(t)})$ for $l = 1,\cdots,m(t)$. Standard calculations then give us $s_{\text{trans}}(t)^2 = \E h_{\text{trans}}(t)^2$ and $s_{\text{PA}}(t)^2 = \E h_{\text{PA}}(t)^2$. Plugging in the estimates $\hat{\bm{A}}$ and $\hat{\bm{B}}$, we can view 
$\hat{s}_{\text{PA}}(t)$ and $\hat{s}_{\text{trans}}(t)$ as the estimates of the expectations of the sample standard deviations in PA and transitivity values observed at the end points of new edges at time-step $t$. As we shall see in Section~\ref{sec:sub_result_contri}, this interpretation also gives us a mean to visualize how well the model fits an observed network.


Finally, we note that this quantification approach is not limited to PA and transitivity. Given a growth formula in which all growth mechanisms are combined in a multiplicative way, for example, as in Eq.~(\ref{eq:probability}), the standard deviations of the logarithmic value of each growth mechanism can be used as a measure of the contribution of that mechanism.

\section{Results and Discussion}\label{sec:result}
\subsection{Two co-authorship networks}\label{sec:sub_result_data}
We applied our proposed method to two different scientific co-authorship networks: SMJ~\cite{pham4} and STA~\cite{stats_dataset}, where nodes represent authors and links represent co-authorship in papers. SMJ includes papers published in the Strategic Management Journal, considered to be one of the top journals in strategy and management, from 1980 to 2017. 
STA includes papers in four statistics journals: the Journal of the American Statistical Association, the Journal of the Royal Statistical Society (Series B), the Annals of Statistics, and Biometrika, from 2003 to 2012. These journals are generally considered the top journals in statistics. New collaborations and repeated collaborations are pooled together in both networks. The time resolution is chosen to be one-year in SMJ and six-month in STA. 

Table~\ref{tb:stats} shows the summary statistics for two networks. The ratios $\Delta |V|/|V|$ and $\Delta |E|/|E|$ are both close to one, which indicate that each network grows out from a very small initial network. Since the number of new edges $\Delta |V|$ is loosely corresponding to the number of available data in our statistical model, STA has the biggest amount of data. The clustering coefficients in both networks are rather high, but nevertheless fall in the normal range observed in real-world networks~\cite{Newman_coauthornet1}.

\begin{table*}[htbp]
\center
 \caption{Summary statistics for two scientific co-authorship networks. $|V|$ and $|E|$ are the total numbers of nodes and edges in the final snapshot, respectively. $T$ is the number of time-steps. $\Delta|V|$ and $\Delta |E|$ are the increments of nodes and edges after the initial snapshot, respectively. $C$ is the clustering coefficient of the final snapshot. $k_{max}$ is the maximum degree and $b_{max}$  is the maximum number of common neighbors.}
  \begin{tabular}{|c|c|c|c|c|c|c|c|c|} \hline
    \ Dataset & $|V|$ & $|E|$ & $T$ &  $\Delta |V|$ & $\Delta |E|$ & $C$ & $k_{max}$ & $b_{max}$ \\ \hline 
     SMJ& 2704 & 4131 & 24 & 1991 & 3538 & 0.378 & 34 & 15 \\
     \ \ \ STA \ \ \ &\ \ 3607 \ \  & \ \ 6808 \ \ & \ \ 20 \ \ & \ \ 3261 \ \ &\ \ 6509 \ \ & \ \ 0.320 \ \ & \ \ 65 \ \ & \ \ 19 \ \ \\ 
\hline
  \end{tabular}
  \center
  \label{tb:stats}
\end{table*}

\begin{figure*}[htbp]
    \centering
        \includegraphics[width=\textwidth]{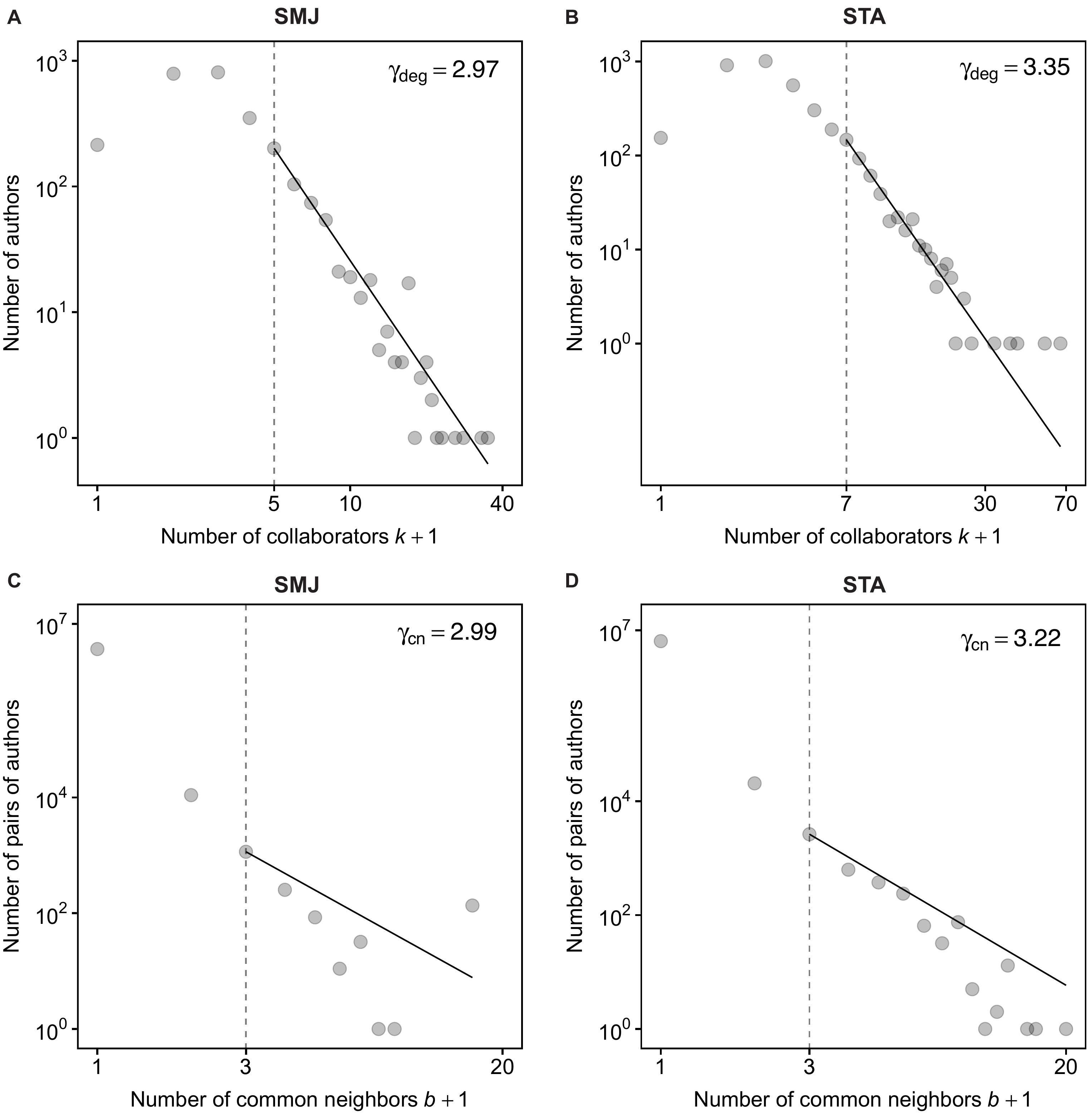}
    \caption{The distributions of $k_i$ and $b_{ij}$ in the final snapshots of two networks. \textbf{A}, \textbf{B}: the degree distributions in the final snapshots of SMJ and STA, respectively. These distributions both display typical shapes of heavy-tails. In each panel, the solid line presents the fitted power-law distribution, and the dotted line indicates where the minimum degree $k_{min}$ is set. \textbf{C}, \textbf{D}: the distributions of the number of pairs with $b$ common neighbors in the final snapshots of SMJ and STA, respectively. In each panel, the solid line presents the fitted power-law distribution, and the dotted line indicates where the minimum number of common neighbours $b_{min}$ is set. In contrast to the degree distributions, the ranges of these distributions of $b_{ij}$ are too narrow for any signs of heavy-tails to emerge.}\label{fig:final_snapshot}
\end{figure*}

It is instructive to look at more fine-grained statistics. Figures~\ref{fig:final_snapshot}A and B show the distributions of the number of collaborators $k$ in the final snapshot of SMJ and STA, respectively. Since the degree distributions in two datasets exhibit signs of heavy-tails, we fitted one of the most representative class of heavy-tail distribution, the power-law distribution $k^{-\gamma_{deg}}$, to these degree distributions by Clauset's procedure~\cite{clauset}. This procedure first chooses the minimum degree $k_{min}$ from which the power-law holds, and then uses a maximum likelihood approach to estimate the power-law exponent $\gamma_{deg}$. The estimated power-law exponents for degree distributions in SMJ and STA are 2.97 and 3.35, respectively. These values fall in the range of $2 < \gamma_{deg} < 4$, which is a commonly observed range for $\gamma_{deg}$ in real-world networks~\cite{newman_powerlaw,clauset}. 

The situation with the distributions of $b_{ij}$ is, however, less clear. Figures~\ref{fig:final_snapshot}C and D show the distributions of the number of node pairs with $b$ common neighbors in the final snapshot of SMJ and STA, respectively. We also fitted the power-law distribution $b^{-\gamma_{cn}}$ to the distributions of $b$ by Clauset's procedure and found that $\gamma_{cn}$ in SMJ and STA are 2.99 and 3.22, respectively. The power-law form, however, seems to be not a very good fit for these distributions. The ranges of $b$ in two distributions seem to be too narrow to say anything definitely about the tails. To the best of our knowledge, no previous work has studied the distributions of $b_{ij}$, either in co-authorship networks or any other network types. Since figuring out the distributional form of $b_{ij}$ is not our main goal, we leave this task as future work.

\subsection{PA and transitivity effects are highly non-power-law} \label{sec:sub_result_estimate}

\begin{figure*}[htbp]
    \centering
        \includegraphics[width=\textwidth]{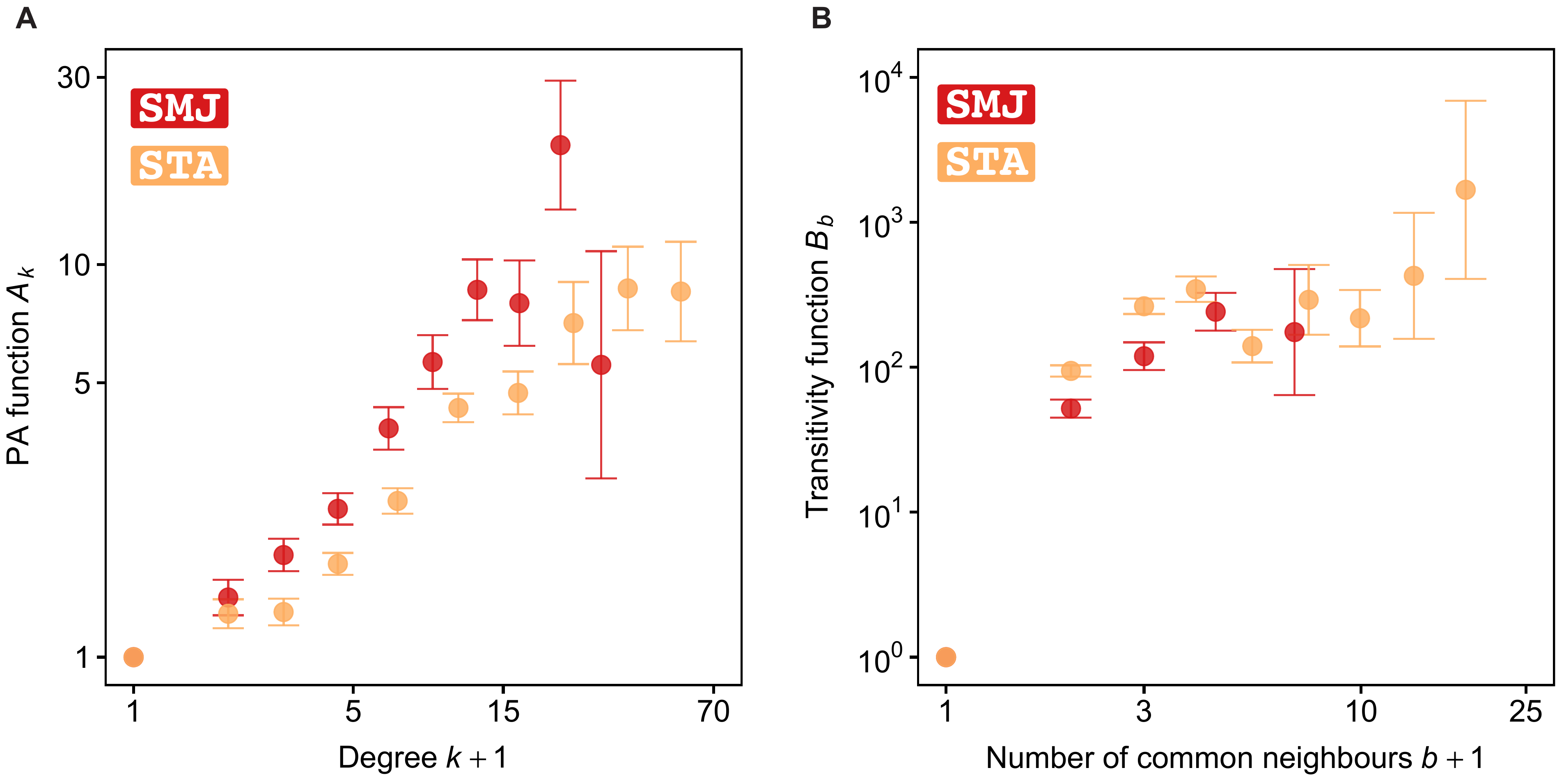}
    \caption{Non-parametric joint estimation of the preferential attachment $A_k$ and transitivity functions $B_b$ in SMJ and STA. The vertical bar at each estimated value indicates the $\pm 2 \sigma$ confidence interval. \textbf{A}: the estimated PA functions are increasing on average in both networks. This implies the existence of the PA phenomenon: a highly-connected author is likely to get more new collaborations than a lowly-connected one. \textbf{B}: The transitivity effect is highly non-power-law in both networks. While $B_b$ greatly increases when $b$ changes from $0$ to $1$ in both networks, after this initial huge jump, $B_b$ stays relatively horizontal in SMJ and only slightly increases in STA. The huge jump at $b = 1$ implies that co-authors of co-authors is at least ten times more likely to become new co-authors, comparing with the case when there is no mutual co-author. }\label{fig:nonpara}
\end{figure*}

Applying the proposed method to two data-sets, we found that the estimated PA and transitivity functions display non-power-law and complex trends (Figures~\ref{fig:nonpara}). 

In both networks, the value of $A_k$ increases on average in $k$, which implies the existence of the PA phenomenon: the more collaborators an author gets, the more likely they would get a new one. This is consistent with previous results in the literature, in which the phenomenon has been found in collaboration networks in diverse fields~\cite{newman2001clustering,miloj,rsiena_slovenian_2012,rsiena_transitivity_1}. 

The situation with the transitivity functions is more complex. In both SMJ and STA, there is a huge jump when $b$ goes from $0$ to $1$: $B_1/B_0$ is about $60$ times in SMJ and almost $100$ times in STA. These jumps in $B_b$ values have been previously observed in co-authorship networks~\cite{newman2001clustering,miloj}. After this initial jump, $B_b$, however, stays relatively horizontal in both SMJ and STA, before slightly increases again in SMJ. This complex departure from the power-law form renders any statement about a universal transitivity effect moot. The value of $B_b$ at every $b > 0$ is, however, at least one order of magnitude higher than $B_0$, which suggests that, co-authors of co-authors seem to be at least ten times more likely to become new co-authors, comparing with the case when there is no mutual co-author.


It is informative to supplement the non-parametric analysis with a parametric one, since the theoretical literature offers many insights in this context. Here, we follow the standard practice and fit the power-law functional forms $A_k = (k+1)^\alpha$ and $B_b = (b+1)^\beta$~\cite{krapi,jeong,pham2}. To find the PA attachment exponent $\alpha$ and the transitivity attachment exponent $\beta$, we substitute these forms to Eq.~(\ref{eq:probability}) and numerically maximizes the resulted log-likelihood function with respected to $\alpha$ and $\beta$. Table~\ref{tb:para} shows the estimated values of $\alpha$ and $\beta$. 

\begin{table*}[htbp]
\center
 \caption{Estimated values of the PA attachment exponent $\alpha$ and the transitivity attachment exponent $\beta$ in two networks. The values are estimated by maximum partial likelihood estimation. The interval given at each estimated value is two-sigma.}
  \begin{tabular}{|c|c|c|c|c|} \hline
    \ Network & PA attachment exponent $\alpha$ & Transitivity attachment exponent $\beta$ \\ \hline 
     SMJ & $0.93\ (\pm 0.04)$ & $2.50\ (\pm 0.07)$ \\
     STA & $0.84\ (\pm 0.03)$ & $3.05\ (\pm 0.04)$ \\
\hline
  \end{tabular}
  \center
  \label{tb:para}
\end{table*}

The PA attachment $\alpha$ in both networks are in the sub-linear region, i.e., $0<\alpha < 1$, which is a frequently observed range in real-world networks~\cite{newman2001clustering,pham2,pham4}. While this region has been shown to give rise to a heavy-tail degree distribution when there is only PA at play~\cite{krapi}, there is no such theoretical result when PA jointly exists with transitivity. It is, however, not entirely unreasonable to expect that the sub-linear value of $\alpha$ is responsible for the observed heavy-tail degree distributions in Figs.~\ref{fig:final_snapshot}A and B.

The transitivity attachment exponents $\beta$ are both greater than $1$, indicating an exponentially faster growth rate of the transitivity function comparing to the PA function. This is evident in, for example, STA: while $A_{10}$ is less than $10$, $B_{10}$ is already larger than $100$. To the best of our knowledge, there is no theoretical result on the effect of $\beta$ on the structure of a growing network, even for the supposedly simpler case when there is only transitivity. 

Overall, the results in this section indicate the joint existence of PA and transitivity phenomena in both networks. Our non-parametric approach revealed that a conventional power-law functional form in a parametric approach may not be the best to describe the two phenomena. For $A_k$, the power-law form fits reasonably well the low-degree part, but cannot capture the deviations from the power-law form in the high-degree part. For $B_b$, the power-law form is even less suitable. We hope our non-parametric findings  could offer hints on more suitable parametric forms for $A_k$ and $B_b$.

\subsection{Transitivity dominated PA in both networks}\label{sec:sub_result_contri}

\begin{figure*}[htbp]
    \centering
        \includegraphics[width=\textwidth]{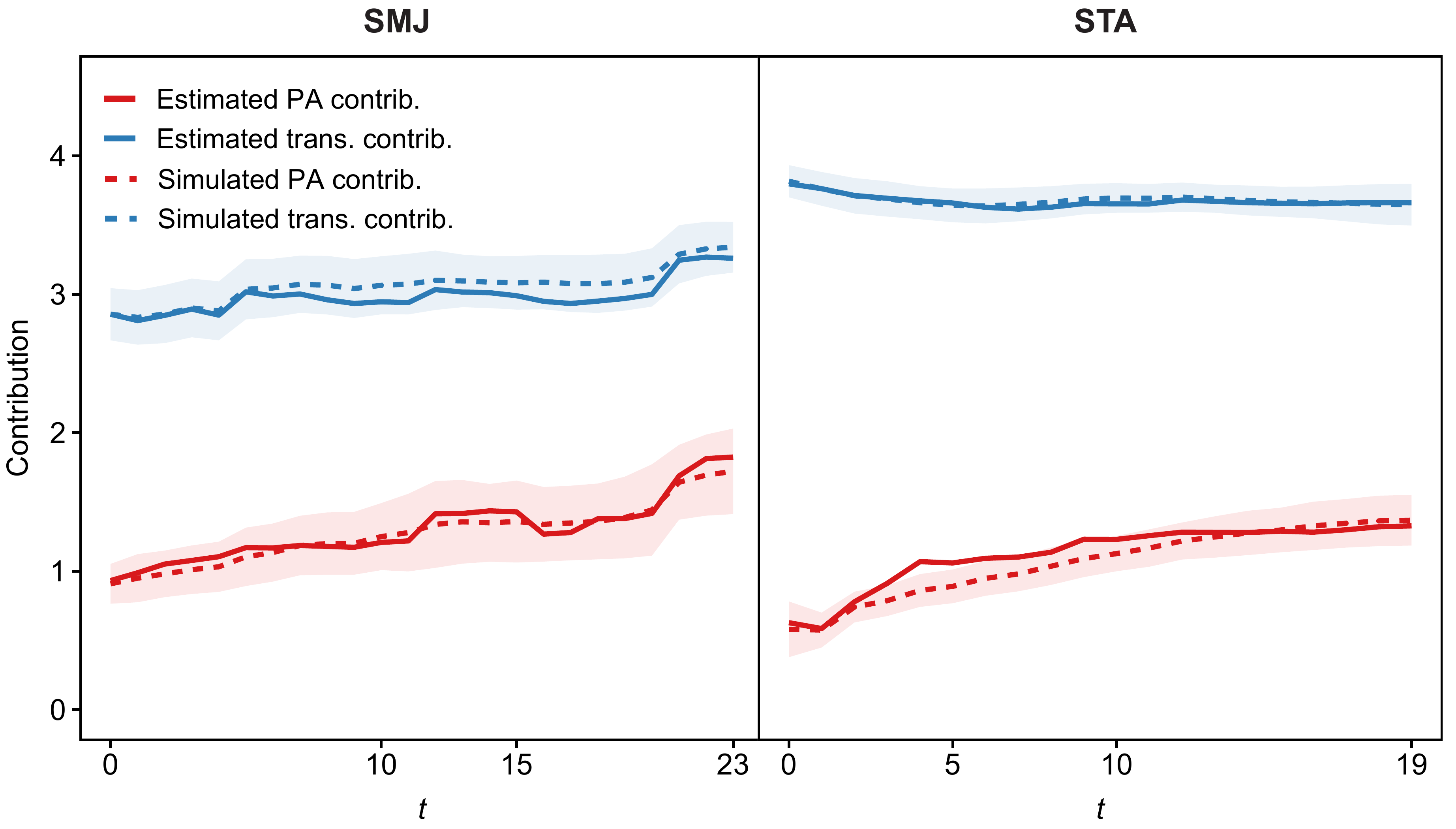}
    \caption{Estimated and simulated contributions of PA and transitivity at each time-step in SMJ and STA. The amount of contribution is measured by $\log_2$ fold-changes in the model Eq.~(\ref{eq:probability}). The solid lines are the estimated contributions $\hat{s}_{\text{PA}}(t)$ and $\hat{s}_{\text{trans}}(t)$ calculated by plugging the estimates $\hat{\bm{A}}$ and $\hat{\bm{B}}$ into Eqs.~(\ref{eq:dynamic_PA}) and~(\ref{eq:dynamic_trans}). Each dotted line is the average of the corresponding true contributions of $100$ simulated networks that use $\hat{\bm{A}}$ and $\hat{\bm{B}}$ as the true functions. The band around each depicts the $\pm$ two times the population standard deviation of the simulated contributions. The solid and dotted lines agree well with each other, which suggests that $\hat{s}_{\text{PA}}(t)$ and $\hat{s}_{\text{trans}}(t)$ are reliable.}\label{fig:variances}
\end{figure*}

After obtaining the estimates $\hat{\mathbf{A}}$ and $\hat{\mathbf{B}}$, we can compute the amount of contributions of PA and transitivity in the growth process of each network by plugging these estimates into Eqs.~(\ref{eq:dynamic_PA}) and~(\ref{eq:dynamic_trans}). The estimated amount of contributions $\hat{s}_{\text{PA}}(t)$ and $\hat{s}_{\text{trans}}(t)$ are shown in Fig.~\ref{fig:variances} as solid lines.

In each network, $\hat{s}_{\text{trans}}(t)$ is greater than $\hat{s}_{\text{PA}}(t)$ for all $t$. One might ask whether these tendencies hold for the true values $s_{\text{PA}}(t)$ and $s_{\text{trans}}(t)$ as well, or they are just artifacts arising when we plug $\hat{\bm{A}}$ and $\hat{\bm{B}}$ into Eqs.~(\ref{eq:dynamic_PA}) and~(\ref{eq:dynamic_trans}). We demonstrate by simulations that, if the true $\bm{A}$ and $\bm{B}$ are close to the estimates $\hat{\bm{A}}$ and $\hat{\bm{B}}$, $s_{\text{PA}}(t)$ and $s_{\text{trans}}(t)$ are similar to $\hat{s}_{\text{PA}}(t)$ and $\hat{s}_{\text{trans}}(t)$. For each real network, we simulated $100$ networks based on Eq.~(\ref{eq:probability}) using the estimates $\hat{\bm{A}}$ and $\hat{\bm{B}}$ as true functions. We kept all the aspects of the growth process that are not governed by Eq.~(\ref{eq:probability}) the same as what observed in the real network. This includes using the observed initial graph and the observed numbers of new nodes and new edges at each time-step in the simulation. Since $\hat{\bm{A}}$ and $\hat{\bm{B}}$ are the true PA and transitivity functions for each simulated network, we were able to calculate the true contributions of PA and transitivity in each simulated network using Eqs.~(\ref{eq:dynamic_PA}) and~(\ref{eq:dynamic_trans}). The behaviours of the simulated contributions are very similar to the estimated contributions $\hat{s}_{\text{PA}}(t)$ and $\hat{s}_{\text{trans}}(t)$, which indicates that the latter are likely to be reliable.

\begin{figure*}[htbp]
    \centering
        \includegraphics[width=\textwidth]{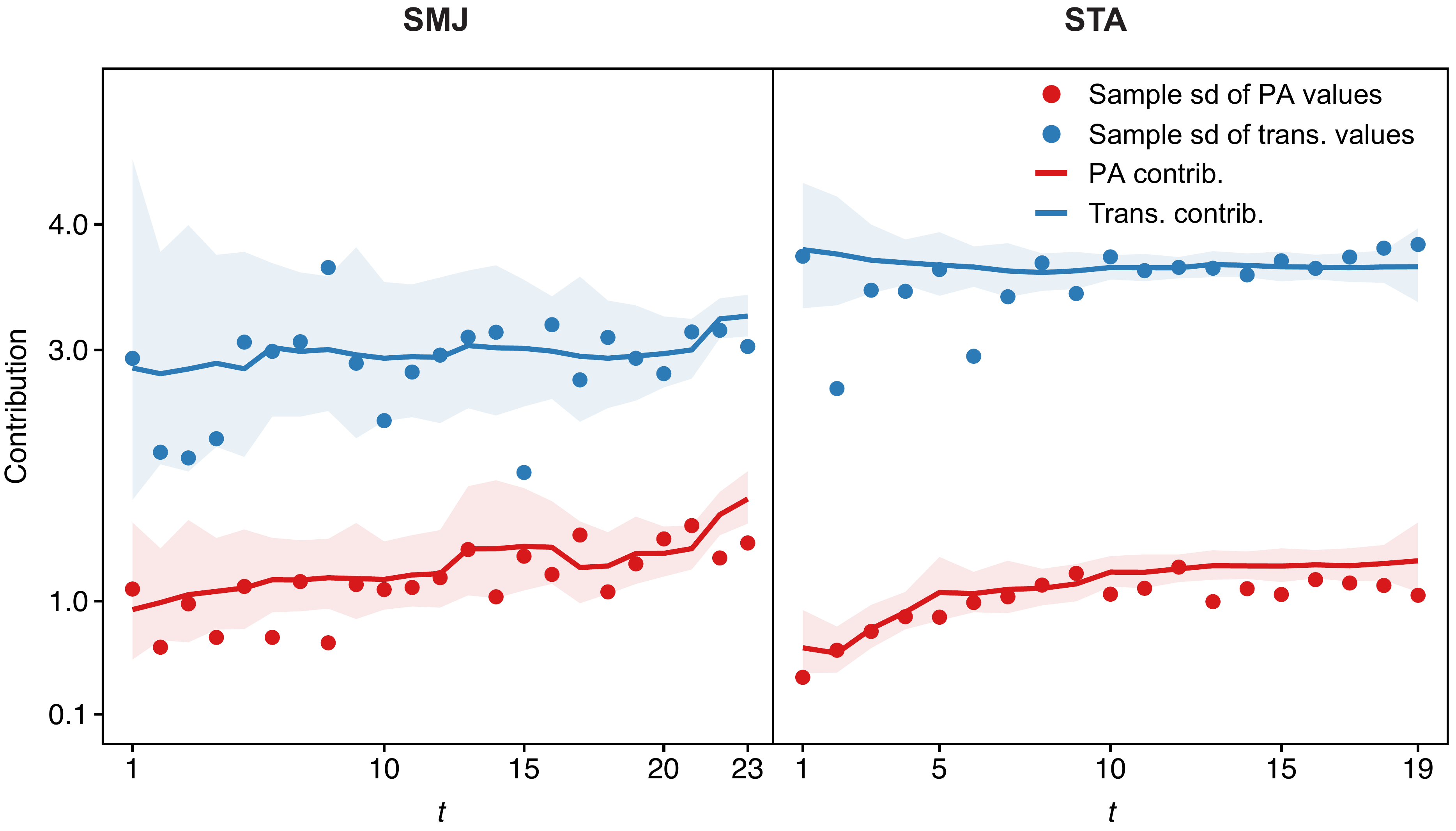}
    \caption{The sample standard deviations of PA and transitivity values at end points of actually-observed new edges, $\hat{h}_{\text{PA}}(t)$ and $\hat{h}_{\text{trans}}(t)$, agree well with their estimated expectations, $\hat{s}_{\text{PA}}(t)$ and $\hat{s}_{\text{trans}}(t)$. This implies that the statistical model is consistent with the observed data. The band around $\hat{s}_{\text{PA}}(t)$ depicts the interval of $\pm$ two standard deviations of $\hat{h}_{\text{PA}}(t)$. The band around $\hat{s}_{\text{trans}}(t)$ is similar.}\label{fig:model_vs_observed_variance}
\end{figure*}

As explained in Section~\ref{sec:quantify}, one can interpret the contributions $\hat{s}_{\text{PA}}(t)$ and $\hat{s}_{\text{trans}}(t)$ as estimates of the expectations of $\hat{h}_{\text{PA}}(t)$ and $\hat{h}_{\text{trans}}(t)$, the sample standard deviations of PA and transitivity values at end points of actually-observed new edges at time-step $t$. This is expressed as
\begin{equation*}
\E \hat{h}_{\text{PA}}(t) \approx \hat{s}_{\text{PA}}(t)  ;\  
\E \hat{h}_{\text{trans}}(t) \approx \hat{s}_{\text{trans}}(t),
\end{equation*}
where the estimates $\hat{s}_{\text{PA}}(t)$ and $\hat{s}_{\text{trans}}(t)$ slightly overestimate the expectations, because
$$\E \hat{h}_{\text{trans}}(t)  \le (\E \hat{h}_{\text{trans}}(t)^2 )^{1/2} \approx \hat{s}_{\text{trans}}(t).
$$
Figure~\ref{fig:model_vs_observed_variance} shows the observed values  $\hat{h}_{\text{PA}}(t)$ and $\hat{h}_{\text{trans}}(t)$, the estimates $\hat{s}_{\text{PA}}(t)$ and $\hat{s}_{\text{trans}}(t)$ of their expectations, and the estimates of their standard deviations (Appendix~\ref{sec:appendix_2}). The observed values generally fall within two standard deviations around the estimates of their expectations, which implies that Eq.~(\ref{eq:probability}) is consistent with the observed data. 

Overall, the data indicate the governing role of transitivity in the growth processes of both networks: it is mostly the differences in the transitivity values that decide where new collaborations are formed. This intuitive result is consistent with previous results which found that common neighbors are more effective than PA at link prediction in co-authorship networks~\cite{transitivity_win_PA_in_link_prediction}. If PA was what dominates, a scientist would only need to indiscriminately acquire as much collaborators as possible in order to boost their number of collaborators in the future. In light of the current result, they, however, might need to be more selective, since a collaborator who has collaborated with a lot of people might offer more advantages.

\subsection{Diagnosis: time-invariance and goodness-of-fit}\label{sec:sub_result_diag}
Finally, we consider two questions that are critical to our real-world data analysis. The first concerns the validity of the time-invariance assumption of $A_k$ and $B_b$ in two networks: in each network, do $A_k$ and $B_b$ stay relatively unchanged throughout the growth process? The second is whether Eq.~(\ref{eq:probability}) is a reasonably good model for the networks. Although Fig.~(\ref{fig:model_vs_observed_variance}) already hinted at an affirmative answer for both questions, we examine each question in finer details.

\subsubsection{Time invariance of the PA and transitivity functions}
One way to answer the first question is to compare the $A_k$ and $B_b$ in Fig.~\ref{fig:nonpara} with the $A_k$ and $B_b$ estimated using only some portion of the growth process, for many different portions. If they are similar, one can conclude that $A_k$ and $B_b$ indeed stay unchanged throughout the growth process, and thus the time-invariance assumption is valid. 

\begin{figure*}[htbp]
    \centering
        \includegraphics[width=\textwidth]{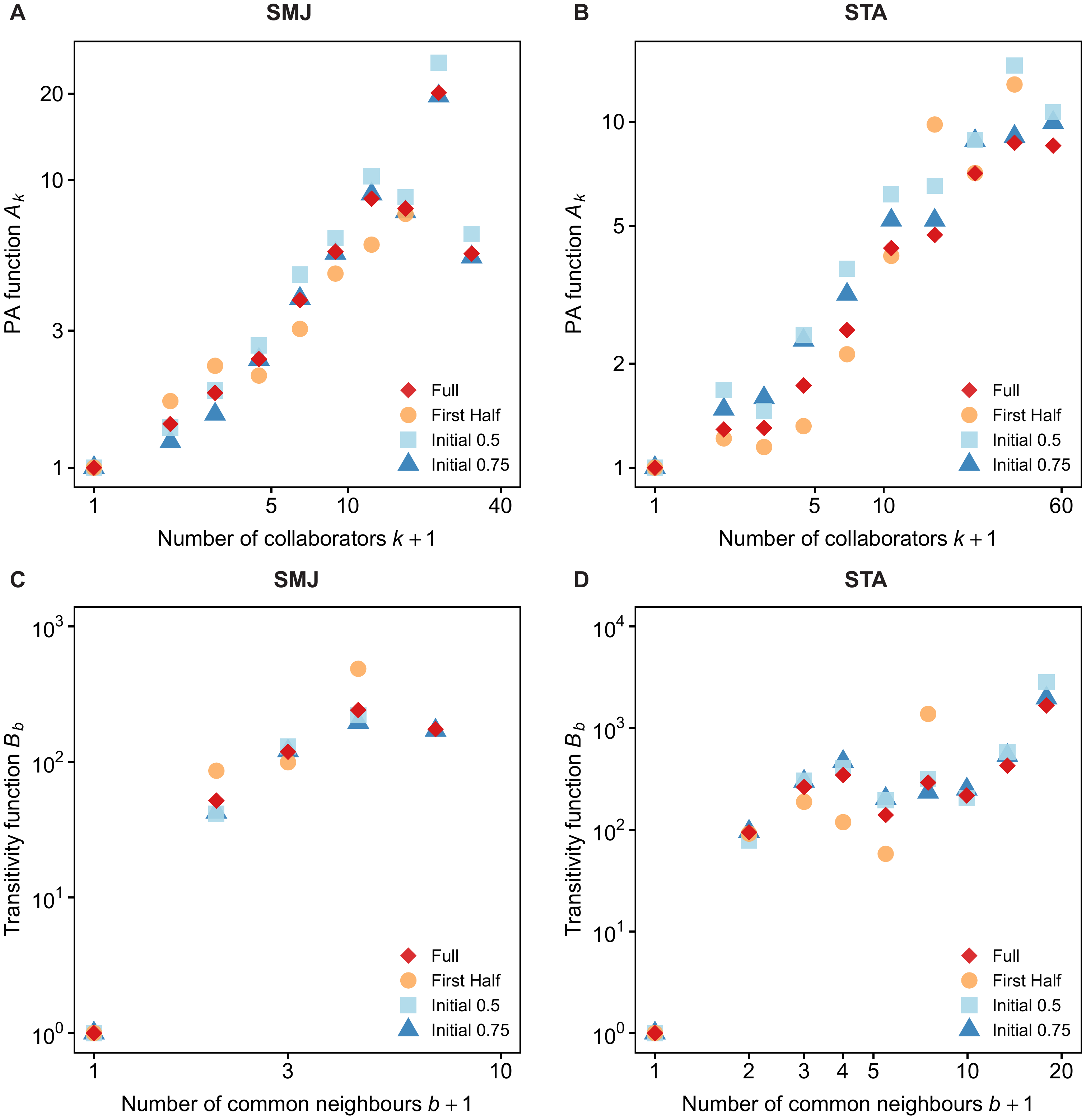}
    
    \caption{Time invariance of the PA and transitivity functions. \textbf{A} and \textbf{C}: PA and transitivity functions of SMJ. \textbf{B} and \textbf{D}: PA and transitivity functions of STA. While ``First Half'' contains only the first half of the growth process, the initial time is set at the middle and at the $3/4$ point of the time-line in  ``Initial 0.5'' and ``Initial 0.75'', respectively. In each data-set, all four PA /transitivity functions agree well with each other, which suggests that the PA and transitivity functions stay relatively unchanged throughout the growth process.}\label{fig:time_invariance}
\end{figure*}

To this end, from each original network, we create three new networks. The first new network (``First Half'') contains only the first half of the growth process, thus allows estimating $A_k$ and $B_b$ in this portion. In the second new network (``Initial 0.5''), we set the initial time at the middle of the time-line, effectively enabling estimation of $A_k$ and $B_b$ of the second half of the growth process. In the third new network (``Initial 0.75''), we set the initial time at the $3/4$ point of the time-line. This network lets us estimating $A_k$ and $B_B$ in the last quarter of the growth process. The estimated $A_k$ and $B_b$ in these three new networks then are compared with the $A_k$ and $B_b$ obtained from the full growth process (Figure~\ref{fig:time_invariance}). Visual inspection of Fig.~\ref{fig:time_invariance} suggests that both the PA and transitivity functions stay relatively unchanged in the growth process of each network. This validates the time-invariance assumption.

\subsubsection{Goodness-of-fit}
We use a simulation-based approach to investigate the goodness-of-fit of the model. For each real-world network, we re-use the simulation data used in Fig.~\ref{fig:variances}, which consists of $100$ simulated networks generated using the estimated $A_k$ and $B_b$ of that network as true functions. We compare some statistics of the simulated networks with the corresponding statistics of the real network. If Eq.~(\ref{eq:probability}) is a good fit, then the observed statistics and the simulated statistics must be close. Similar simulation-based approaches have been proposed for inspecting goodness-of-fit of exponential random-graph models~\cite{ergm_gof} and stochastic actor-based models~\cite{stochastic_actor_gof_1,stochastic_actor_phd}. 

\begin{figure*}[htbp]
    \centering
        \includegraphics[width=\textwidth]{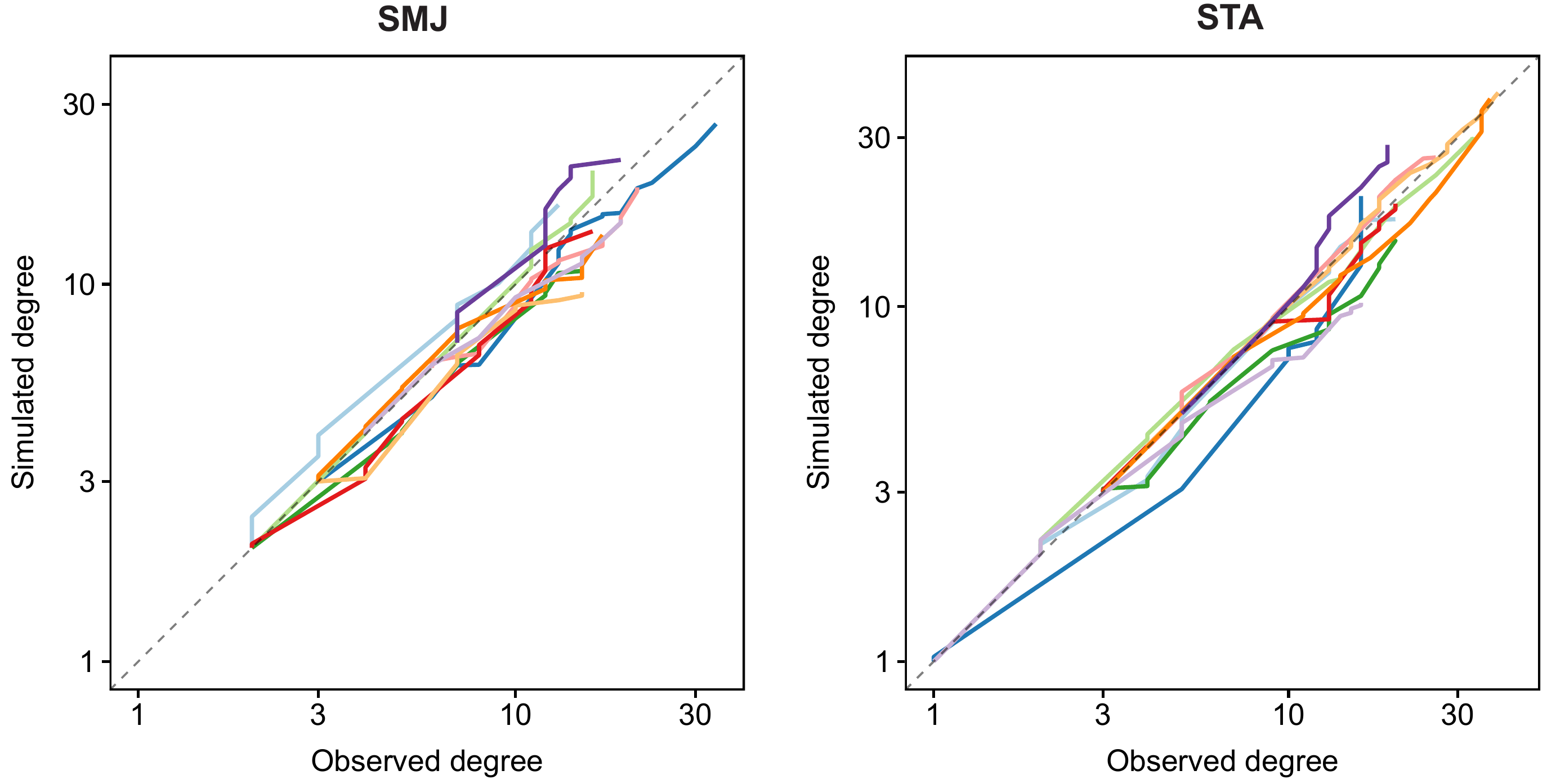}
    \caption{Pairs of observed and simulated degree of some high-degree nodes in two networks. The simulation data is the same as the simulation data used in Fig.~\ref{fig:variances}. In each network, ten nodes are chosen uniformly at random from the top $1\%$ of all nodes in term of the number of new edges accumulated during the growth process. Each line represents the observed degrees and the corresponding simulated values at each time-step of a node. Each simulated value is averaged over $100$ simulations. In each network, the pairs of observed and simulated degrees are reasonably close to the identity line, which suggests the model fits the degree curves well.}\label{fig:growth_curve}
\end{figure*}

For an overview, we look at how well the model can replicate the observed degree curves. In Fig.~\ref{fig:growth_curve}, for each real-world network we choose uniformly at random ten nodes from the top $1\%$ of all nodes in term of the number of new edges accumulated during the growth process. For each node, we then plot the evolution line of the observed degree value and the simulated degree value. The closer this line to the line of equality is, the better the model captures the observed degree growth of that node. Although for some nodes the simulated degree sometimes tends to be lower than the observed degree, the lines are overall reasonably close to the identity line, which implies the model captured the degree growth well. 

For a closer inspection, we then look at how well the model replicates the probability distribution of new edges during the growth process. In particular, consider sampling uniformly at random an edge $e$ from the set of all new edges in the growth process. Suppose that $e$ is between a node pair with degrees $K_1$ and $K_2$ ($K_1 \le K_2$) and the number of their common neighbors is $X$. The relative frequency, or observed probability, that $K_1 = k_1$, $K_2 = k_2$, and $X = b$ is 
$p_{k_1,k_2,b} = \sum_{t}m_{k_1,k_2,b}(t)/\sum_{k_1=0}^{k_{max}}\sum_{k_2=k_1}^{k_{max}}\sum_{b=0}^{b_{max}}\sum_{t}m_{k_1,k_2,b}(t)$,
in which $m_{k_1,k_2,b}(t)$ is the number of new edges emerged at time $t$ between a node pair whose degrees are $k_1$ and $k_2$ and their number of common neighbors is $b$. The probability $p_{k_1,k_2,b}$ thus summarizes information about the associations of $k_1$, $k_2$, and $b$ at the end points of new edges through out the growth process. 
Our joint estimation of PA and transitivity is compared with two conventional approaches in which PA~\cite{pham2} and transitivity~\cite{newman2001clustering} are estimated in isolation. For each of these two approaches, we first estimate the PA/transitivity function in isolation and then use the estimated function to generate $100$ networks in order to inspect how well each existing method replicates $p_{k_1,k_2,b}$. In order to visualize this probability distribution, which is multi-dimensional, we slice it into many one-dimensional ones by conditioning. 

\begin{figure*}[htbp]
    \centering
        \includegraphics[width=\textwidth]{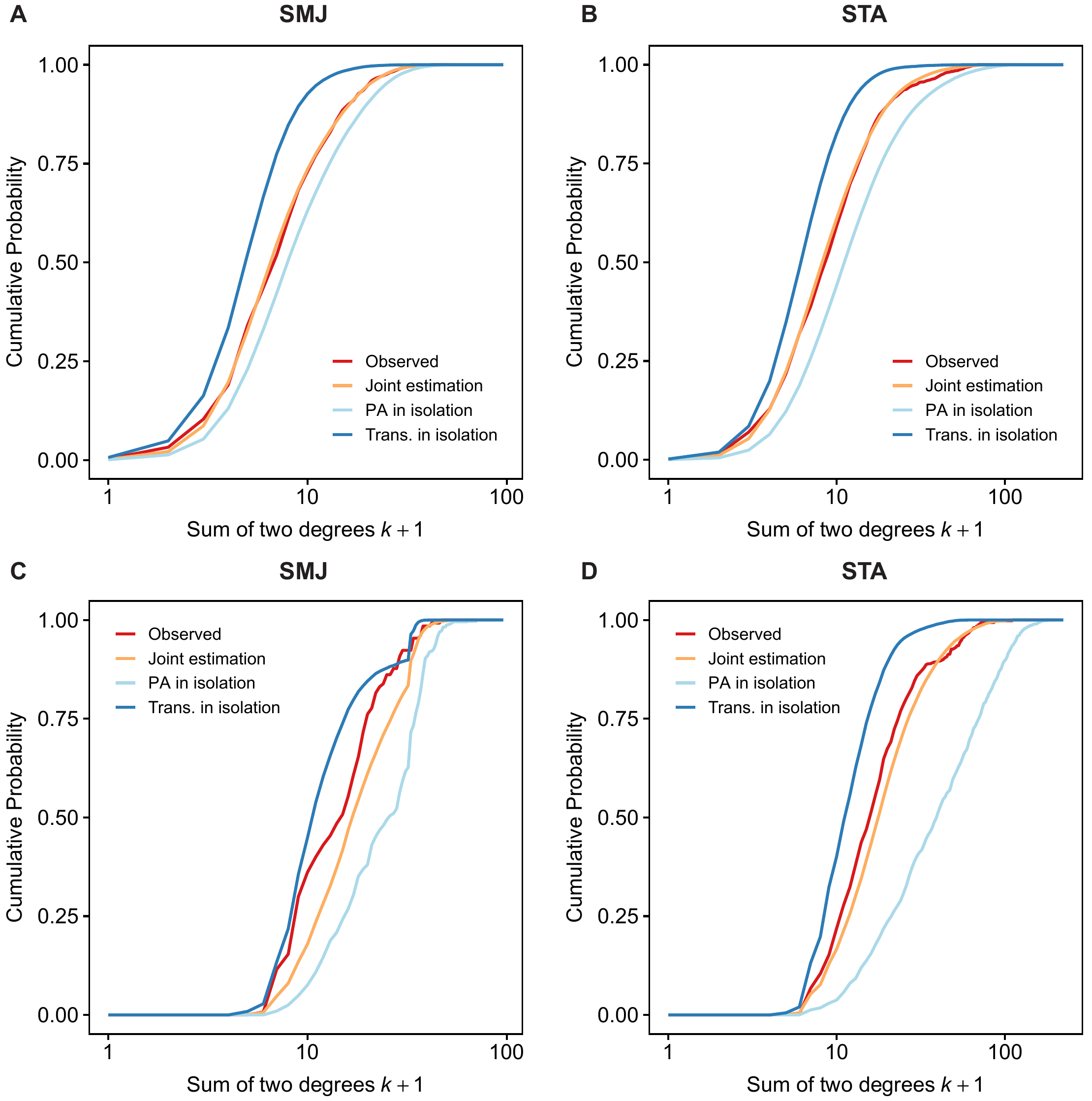}
    \caption{Observed and simulated cumulative probability distributions $p_{k|b \le 1}$ and $p_{k|b \ge 2}$ of $k = k_1 + k_2$ in two networks. For each estimation method, we generate $100$ networks from the estimation result and report the average values over $100$ simulations. \textbf{A} and \textbf{B}:  the cumulative probability distribution $p_{k|b \le 1}$ in SMJ and STA, respectively. \textbf{C} and \textbf{D}: the cumulative probability distribution $p_{k|b \ge 2}$ in SMJ and STA, respectively. In all cases, our joint estimation approach replicated the observed distributions comparatively well.}\label{fig:k_dist_reconstruct}
\end{figure*}

Firstly, we look at 
$$p_{k|b \in \mathcal{B}} := Pr(K_1 + K_2 = k | X\in \mathcal{B}) = \sum_{b\in \mathcal{B}}\sum_{k_1 = 0}^{k_{max}}p_{k_1,k-k_1,b}/\sum_{b \in \mathcal{B}}\sum_{k_1=0}^{k_{max}}\sum_{k_2 = k_1}^{k_{max}}p_{k_1,k_2,b},$$ with the convention that $p_{k_1,k_2,b} = 0$ whenever $k_1 > k_2$ or $k_2 > k_{max}$. This is the probability distribution of $K_1 + K_2$ conditioning on the event $X \in \mathcal{B}$. Since we know from Fig.~\ref{fig:final_snapshot} that the number of node pairs with $b = 0$ or $b = 1$ is vastly greater than the rest, we consider two probability distributions $p_{k|b \le 1}$ and $p_{k|b \ge 2}$ and show their cumulative probability distributions in Fig.~\ref{fig:k_dist_reconstruct}. In all cases, the joint estimation approach best replicated the observed distributions. It is surprising to observe that the $B_b$-in-isolation approach, which does not explicitly leverage any information about $k$, has more or less the same replication performance as the $A_k$-in-isolation approach, which explicitly does. This suggests that the dimension of $b$ preserves a fair amount of the information about $k$.

\begin{figure*}[htbp]
    \centering
      \includegraphics[width=\textwidth]{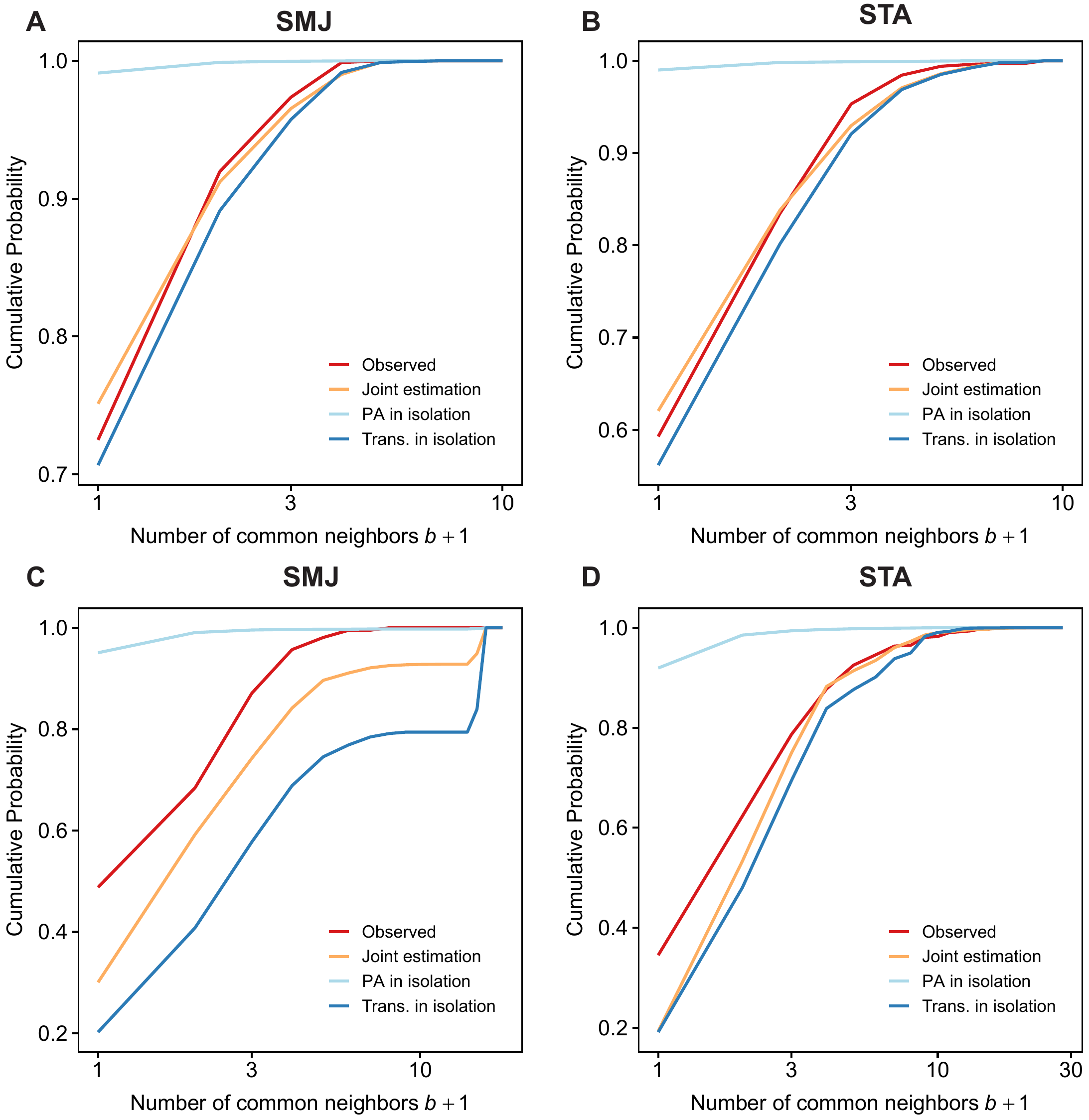}
    \caption{Observed and simulated cumulative probability distributions $p_{b|\max(k_1,k_2) \le 9}$ and $p_{b|\max (k_1,k_2) \ge 10}$ in two networks. For each estimation method, we generate $100$ networks from the estimation result and report the average values over $100$ simulations. \textbf{A} and \textbf{B}: the cumulative probability distribution $p_{b|\max(k_1,k_2) \le 9}$ in SMJ and STA, respectively. \textbf{C} and \textbf{D}: the cumulative probability distribution of $p_{b|\max (k_1,k_2) \ge 10}$ in SMJ and STA, respectively. In all cases, our joint estimation approach replicated the observed distributions comparatively well.}\label{fig:b_dist_reconstruct}
\end{figure*}

Secondly, we look at 
$$
p_{b|(k_1,k_2) \in \mathcal{K}} := Pr(X = b | (K_1,K_2)\in \mathcal{K}) = \sum_{(k_1,k_2)\in \mathcal{K}}p_{k_1,k_2,b}/\sum_{b=0}^{b_{max}}\sum_{(k_1,k_2)\in \mathcal{K}}p_{k_1,k_2,b},$$
where $\mathcal{K}$ is a non-empty set of un-ordered pairs. This is the probability distribution of $X$ conditioning on the event $(K_1,K_2) \in \mathcal{K}$. Given a pair of node whose degrees are $k_1$ and $k_2$ and their number of common neighbours is $b$, there is a natural condition imposed on $b$: $b$ must be not greater than either $k_1$ or $k_2$. So if one chooses $\mathcal{K}$ such that $k_1$ or $k_2$ could be too small, the range of $b$ would be severely limited. For this reason, we consider two probability distributions: $p_{b|\max(k_1,k_2) \le 9}$ and $p_{b|\max (k_1,k_2) \ge 10}$, both allow a large range for $b$. Their cumulative distributions are shown in Fig.~\ref{fig:b_dist_reconstruct}. Once again, the joint estimation approach best replicated the observed cumulative probability distributions in all cases. While the $B_b$-in-isolation approach replicated fairly well the observed distributions in most cases, the $A_k$-in-isolation approach completely failed to do so in all cases. This implies that, while the dimension of $b$ seems to preserve a fair amount of the information about $k_1$ and $k_2$, the dimensions of $k_1$ and $k_2$ maintain little information about $b$.

Overall, the joint estimation approach performed comparatively well. The surprisingly good performance of the $B_b$-in-isolation approach is, in fact, in agreement with the dominating role of $B_b$ in the growth process of both networks. Combining the results in Fig.~\ref{fig:growth_curve} with those in Figs.~\ref{fig:k_dist_reconstruct} and \ref{fig:b_dist_reconstruct}, we conclude that the joint estimation approach captured reasonably well both first-order and second-order information of the networks. This good fit is consistent with the fact that the key assumption of time-invariability of $A_k$ and $B_b$ is satisfied in both networks.

\section{Conclusion}\label{sec:conclusion}

We proposed a statistical network model that incorporates non-parametric PA and transitivity functions and derived an efficient MM algorithm for estimating its parameters. We also presented a method that is able to quantify the amount of contributions of not only PA and transitivity but also many other network growth mechanisms by exploiting the probabilistic dynamic process induced by the model formula.

We showed that the proposed network model is a reasonably good fit to two real-world co-authorship networks and revealed intriguing properties of the PA and transitivity functions in those networks. The PA function is increasing on average in both networks, which implies the PA effect is at play. Excluding the high degree part, it does follow the conventional power-law form reasonably well. The transitivity function is, however, highly non-power-law in two networks: it jumps greatly after $b =0$, but stays relatively horizontal or only slightly increases afterwards. This non-conventional form implies that co-authors of co-authors seems to be at least ten times more likely to become new co-authors, comparing with the case when there is no mutual co-author. We also found transitivity dominating PA in both networks, which suggests the importance of indirect relations in scientific creative processes.

There are some fascinating directions for further developing the statistical methodology. Firstly, although the proposed model and most other network models in the literature assume that new edges at each time-step are independent, such edges are hardly so in real-world collaboration networks. Efficiently relaxing this assumption might lead to better models for this network type. Secondly, it is curious to see whether one could take the time-invariability test developed for stochastic actor-based models~\cite{stochastic_actor_model_test_time} and adapt it to our model. 

On the application front, this work lays out a potentially fruitful approach for analyzing complex networks, while raising more questions than it answers. Does transitivity always dominate PA in co-authorship networks? Which parametric forms are capable of capturing the fine details seen in Fig.~\ref{fig:nonpara}? What are the properties of PA and transitivity in co-authorship networks at the level of institutions or countries? We hope this paper has convinced informetricians to include non-parametric modelling of PA and transitivity into their toolbox.

\section{Acknowledgements}
This work was supported in part by JSPS KAKENHI Grant Numbers JP19K20231 to TP and JP16H02789 to HS. The funding source had no role in study design; in the collection, analysis and interpretation of data; in the writing of the report; and in the decision to submit the article for publication.

\bibliographystyle{apacite}
\bibliography{library.bib}

\begin{thebibliography}{}

\bibitem [\protect \citeauthoryear {%
Albert%
\ \BBA {} Barab{\'a}si%
}{%
Albert%
\ \BBA {} Barab{\'a}si%
}{%
{\protect \APACyear {1999}}%
}]{%
barabasi-albert}
\APACinsertmetastar {%
barabasi-albert}%
\begin{APACrefauthors}%
Albert, R.%
\BCBT {}\ \BBA {} Barab{\'a}si, A.%
\end{APACrefauthors}%
\unskip\
\newblock
\APACrefYearMonthDay{1999}{}{}.
\newblock
{\BBOQ}\APACrefatitle {Emergence of Scaling in Random Networks} {Emergence of
  scaling in random networks}.{\BBCQ}
\newblock
\APACjournalVolNumPages{Science}{286}{}{509--512}.
\newblock
\begin{APACrefURL} \url{https://doi.org/10.1126/science.286.5439.509}
  \end{APACrefURL}
\PrintBackRefs{\CurrentBib}

\bibitem [\protect \citeauthoryear {%
Bornmann%
}{%
Bornmann%
}{%
{\protect \APACyear {2017}}%
}]{%
citation_2}
\APACinsertmetastar {%
citation_2}%
\begin{APACrefauthors}%
Bornmann, L.%
\end{APACrefauthors}%
\unskip\
\newblock
\APACrefYearMonthDay{2017}{}{}.
\newblock
{\BBOQ}\APACrefatitle {Is Collaboration Among Scientists Related to the
  Citation Impact of Papers Because Their Quality Increases with Collaboration?
  {A}n Analysis Based on Data from F1000Prime and Normalized Citation Scores}
  {Is collaboration among scientists related to the citation impact of papers
  because their quality increases with collaboration? {A}n analysis based on
  data from f1000prime and normalized citation scores}.{\BBCQ}
\newblock
\APACjournalVolNumPages{Journal of the Association for Information Science and
  Technology}{68}{4}{1036--1047}.
\newblock
\begin{APACrefURL} \url{https://doi.org/10.1002/asi.23728} \end{APACrefURL}
\PrintBackRefs{\CurrentBib}

\bibitem [\protect \citeauthoryear {%
Callaway%
, Hopcroft%
, Kleinberg%
, Newman%
\BCBL {}\ \BBA {} Strogatz%
}{%
Callaway%
\ \protect \BOthers {.}}{%
{\protect \APACyear {2001}}%
}]{%
grow-random}
\APACinsertmetastar {%
grow-random}%
\begin{APACrefauthors}%
Callaway, D\BPBI S.%
, Hopcroft, J\BPBI E.%
, Kleinberg, J\BPBI M.%
, Newman, M\BPBI E\BPBI J.%
\BCBL {}\ \BBA {} Strogatz, S\BPBI H.%
\end{APACrefauthors}%
\unskip\
\newblock
\APACrefYearMonthDay{2001}{}{}.
\newblock
{\BBOQ}\APACrefatitle {Are randomly grown graphs really random?} {Are randomly
  grown graphs really random?}{\BBCQ}
\newblock
\APACjournalVolNumPages{Physical Review E}{64}{}{041902}.
\newblock
\begin{APACrefURL} \url{https://doi.org/10.1103/PhysRevE.64.041902}
  \end{APACrefURL}
\PrintBackRefs{\CurrentBib}

\bibitem [\protect \citeauthoryear {%
Clauset%
, Shalizi%
\BCBL {}\ \BBA {} Newman%
}{%
Clauset%
\ \protect \BOthers {.}}{%
{\protect \APACyear {2009}}%
}]{%
clauset}
\APACinsertmetastar {%
clauset}%
\begin{APACrefauthors}%
Clauset, A.%
, Shalizi, C\BPBI R.%
\BCBL {}\ \BBA {} Newman, M\BPBI E\BPBI J.%
\end{APACrefauthors}%
\unskip\
\newblock
\APACrefYearMonthDay{2009}{}{}.
\newblock
{\BBOQ}\APACrefatitle {Power-Law Distributions in Empirical Data} {Power-law
  distributions in empirical data}.{\BBCQ}
\newblock
\APACjournalVolNumPages{SIAM Review}{51}{4}{661--703}.
\newblock
\begin{APACrefURL} \url{https://doi.org/10.1137/070710111} \end{APACrefURL}
\PrintBackRefs{\CurrentBib}

\bibitem [\protect \citeauthoryear {%
Conaldi%
, Lomi%
\BCBL {}\ \BBA {} Tonellato%
}{%
Conaldi%
\ \protect \BOthers {.}}{%
{\protect \APACyear {2012}}%
}]{%
stochastic_actor_gof_1}
\APACinsertmetastar {%
stochastic_actor_gof_1}%
\begin{APACrefauthors}%
Conaldi, G.%
, Lomi, A.%
\BCBL {}\ \BBA {} Tonellato, M.%
\end{APACrefauthors}%
\unskip\
\newblock
\APACrefYearMonthDay{2012}{}{}.
\newblock
{\BBOQ}\APACrefatitle {Dynamic Models of Affiliation and the Network Structure
  of Problem Solving in an Open Source Software Project} {Dynamic models of
  affiliation and the network structure of problem solving in an open source
  software project}.{\BBCQ}
\newblock
\APACjournalVolNumPages{Organizational Research Methods}{15}{3}{385--412}.
\newblock
\begin{APACrefURL} \url{https://doi.org/10.1177/1094428111430541}
  \end{APACrefURL}
\PrintBackRefs{\CurrentBib}

\bibitem [\protect \citeauthoryear {%
Cs{\'a}rdi%
, Strandburg%
, Zal{\'a}nyi%
, Tobochnik%
\BCBL {}\ \BBA {} {\'E}rdi%
}{%
Cs{\'a}rdi%
\ \protect \BOthers {.}}{%
{\protect \APACyear {2007}}%
}]{%
gabor}
\APACinsertmetastar {%
gabor}%
\begin{APACrefauthors}%
Cs{\'a}rdi, G.%
, Strandburg, K\BPBI J.%
, Zal{\'a}nyi, L.%
, Tobochnik, J.%
\BCBL {}\ \BBA {} {\'E}rdi, P.%
\end{APACrefauthors}%
\unskip\
\newblock
\APACrefYearMonthDay{2007}{}{}.
\newblock
{\BBOQ}\APACrefatitle {Modeling innovation by a kinetic description of the
  patent citation system} {Modeling innovation by a kinetic description of the
  patent citation system}.{\BBCQ}
\newblock
\APACjournalVolNumPages{Physica A: Statistical Mechanics and its
  Applications}{374}{2}{783--793}.
\newblock
\begin{APACrefURL} \url{https://doi.org/10.1016/j.physa.2006.08.022}
  \end{APACrefURL}
\PrintBackRefs{\CurrentBib}

\bibitem [\protect \citeauthoryear {%
Ferligoj%
, Kronegger%
, Mali%
, Snijders%
\BCBL {}\ \BBA {} Doreian%
}{%
Ferligoj%
\ \protect \BOthers {.}}{%
{\protect \APACyear {2015}}%
}]{%
rsiena_transitivity_1}
\APACinsertmetastar {%
rsiena_transitivity_1}%
\begin{APACrefauthors}%
Ferligoj, A.%
, Kronegger, L.%
, Mali, F.%
, Snijders, T\BPBI A\BPBI B.%
\BCBL {}\ \BBA {} Doreian, P.%
\end{APACrefauthors}%
\unskip\
\newblock
\APACrefYearMonthDay{2015}{}{}.
\newblock
{\BBOQ}\APACrefatitle {Scientific collaboration dynamics in a national
  scientific system} {Scientific collaboration dynamics in a national
  scientific system}.{\BBCQ}
\newblock
\APACjournalVolNumPages{Scientometrics}{104}{3}{985--1012}.
\newblock
\begin{APACrefURL} \url{https://doi.org/10.1007/s11192-015-1585-7}
  \end{APACrefURL}
\PrintBackRefs{\CurrentBib}

\bibitem [\protect \citeauthoryear {%
Fortunato%
\ \protect \BOthers {.}}{%
Fortunato%
\ \protect \BOthers {.}}{%
{\protect \APACyear {2018}}%
}]{%
scisci1}
\APACinsertmetastar {%
scisci1}%
\begin{APACrefauthors}%
Fortunato, S.%
, Bergstrom, C\BPBI T.%
, B{\"o}rner, K.%
, Evans, J\BPBI A.%
, Helbing, D.%
, Milojevi{\'c}, S.%
\BDBL {}Barab{\'a}si, A\BHBI L.%
\end{APACrefauthors}%
\unskip\
\newblock
\APACrefYearMonthDay{2018}{}{}.
\newblock
{\BBOQ}\APACrefatitle {Science of science} {Science of science}.{\BBCQ}
\newblock
\APACjournalVolNumPages{Science}{359}{6379}{}.
\newblock
\begin{APACrefURL} \url{https://doi.org/10.1126/science.aao0185}
  \end{APACrefURL}
\PrintBackRefs{\CurrentBib}

\bibitem [\protect \citeauthoryear {%
G\'{o}mez%
, Kappen%
\BCBL {}\ \BBA {} Kaltenbrunner%
}{%
G\'{o}mez%
\ \protect \BOthers {.}}{%
{\protect \APACyear {2011}}%
}]{%
Gomez}
\APACinsertmetastar {%
Gomez}%
\begin{APACrefauthors}%
G\'{o}mez, V.%
, Kappen, H\BPBI J.%
\BCBL {}\ \BBA {} Kaltenbrunner, A.%
\end{APACrefauthors}%
\unskip\
\newblock
\APACrefYearMonthDay{2011}{}{}.
\newblock
{\BBOQ}\APACrefatitle {Modeling the Structure and Evolution of Discussion
  Cascades} {Modeling the structure and evolution of discussion
  cascades}.{\BBCQ}
\newblock
\BIn{} \APACrefbtitle {Proceedings of the 22Nd {ACM} Conference on Hypertext
  and Hypermedia} {Proceedings of the 22nd {ACM} conference on hypertext and
  hypermedia}\ (\BPGS\ 181--190).
\newblock
\APACaddressPublisher{New York, NY, USA}{ACM}.
\newblock
\begin{APACrefURL} \url{https://doi.org/10.1145/1995966.1995992}
  \end{APACrefURL}
\PrintBackRefs{\CurrentBib}

\bibitem [\protect \citeauthoryear {%
Heider%
}{%
Heider%
}{%
{\protect \APACyear {1946}}%
}]{%
fritz}
\APACinsertmetastar {%
fritz}%
\begin{APACrefauthors}%
Heider, F.%
\end{APACrefauthors}%
\unskip\
\newblock
\APACrefYearMonthDay{1946}{}{}.
\newblock
{\BBOQ}\APACrefatitle {Attitudes and Cognitive Organization} {Attitudes and
  cognitive organization}.{\BBCQ}
\newblock
\APACjournalVolNumPages{The Journal of Psychology}{21}{1}{107--112}.
\newblock
\begin{APACrefURL} \url{https://doi.org/10.1080/00223980.1946.9917275}
  \end{APACrefURL}
\PrintBackRefs{\CurrentBib}

\bibitem [\protect \citeauthoryear {%
Holland%
\ \BBA {} Leinhardt%
}{%
Holland%
\ \BBA {} Leinhardt%
}{%
{\protect \APACyear {1970}}%
}]{%
holland_1970}
\APACinsertmetastar {%
holland_1970}%
\begin{APACrefauthors}%
Holland, P\BPBI W.%
\BCBT {}\ \BBA {} Leinhardt, S.%
\end{APACrefauthors}%
\unskip\
\newblock
\APACrefYearMonthDay{1970}{}{}.
\newblock
{\BBOQ}\APACrefatitle {A Method for Detecting Structure in Sociometric Data} {A
  method for detecting structure in sociometric data}.{\BBCQ}
\newblock
\APACjournalVolNumPages{American Journal of Sociology}{76}{3}{492--513}.
\newblock
\begin{APACrefURL} \url{https://doi.org/10.1086/224954} \end{APACrefURL}
\PrintBackRefs{\CurrentBib}

\bibitem [\protect \citeauthoryear {%
Holland%
\ \BBA {} Leinhardt%
}{%
Holland%
\ \BBA {} Leinhardt%
}{%
{\protect \APACyear {1971}}%
}]{%
holland_1971}
\APACinsertmetastar {%
holland_1971}%
\begin{APACrefauthors}%
Holland, P\BPBI W.%
\BCBT {}\ \BBA {} Leinhardt, S.%
\end{APACrefauthors}%
\unskip\
\newblock
\APACrefYearMonthDay{1971}{}{}.
\newblock
{\BBOQ}\APACrefatitle {Transitivity in Structural Models of Small Groups}
  {Transitivity in structural models of small groups}.{\BBCQ}
\newblock
\APACjournalVolNumPages{Comparative Group Studies}{2}{2}{107--124}.
\newblock
\begin{APACrefURL} \url{https://doi.org/10.1177/104649647100200201}
  \end{APACrefURL}
\PrintBackRefs{\CurrentBib}

\bibitem [\protect \citeauthoryear {%
Holland%
\ \BBA {} Leinhardt%
}{%
Holland%
\ \BBA {} Leinhardt%
}{%
{\protect \APACyear {1976}}%
}]{%
holland_1975}
\APACinsertmetastar {%
holland_1975}%
\begin{APACrefauthors}%
Holland, P\BPBI W.%
\BCBT {}\ \BBA {} Leinhardt, S.%
\end{APACrefauthors}%
\unskip\
\newblock
\APACrefYearMonthDay{1976}{}{}.
\newblock
{\BBOQ}\APACrefatitle {Local Structure in Social Networks} {Local structure in
  social networks}.{\BBCQ}
\newblock
\APACjournalVolNumPages{Sociological Methodology}{7}{}{1--45}.
\newblock
\begin{APACrefURL} \url{https://doi.org/10.2307/270703} \end{APACrefURL}
\PrintBackRefs{\CurrentBib}

\bibitem [\protect \citeauthoryear {%
Holland%
\ \BBA {} Leinhardt%
}{%
Holland%
\ \BBA {} Leinhardt%
}{%
{\protect \APACyear {1977}}%
}]{%
holland_1977}
\APACinsertmetastar {%
holland_1977}%
\begin{APACrefauthors}%
Holland, P\BPBI W.%
\BCBT {}\ \BBA {} Leinhardt, S.%
\end{APACrefauthors}%
\unskip\
\newblock
\APACrefYearMonthDay{1977}{}{}.
\newblock
{\BBOQ}\APACrefatitle {A dynamic model for social networks} {A dynamic model
  for social networks}.{\BBCQ}
\newblock
\APACjournalVolNumPages{The Journal of Mathematical Sociology}{5}{1}{5--20}.
\newblock
\begin{APACrefURL} \url{https://doi.org/10.1080/0022250X.1977.9989862}
  \end{APACrefURL}
\PrintBackRefs{\CurrentBib}

\bibitem [\protect \citeauthoryear {%
Hunter%
, Goodreau%
\BCBL {}\ \BBA {} Handcock%
}{%
Hunter%
\ \protect \BOthers {.}}{%
{\protect \APACyear {2008}}%
}]{%
ergm_gof}
\APACinsertmetastar {%
ergm_gof}%
\begin{APACrefauthors}%
Hunter, D\BPBI R.%
, Goodreau, S\BPBI M.%
\BCBL {}\ \BBA {} Handcock, M\BPBI S.%
\end{APACrefauthors}%
\unskip\
\newblock
\APACrefYearMonthDay{2008}{}{}.
\newblock
{\BBOQ}\APACrefatitle {Goodness of Fit of Social Network Models} {Goodness of
  fit of social network models}.{\BBCQ}
\newblock
\APACjournalVolNumPages{Journal of the American Statistical
  Association}{103}{481}{248--258}.
\newblock
\begin{APACrefURL} \url{https://doi.org/10.1198/016214507000000446}
  \end{APACrefURL}
\PrintBackRefs{\CurrentBib}

\bibitem [\protect \citeauthoryear {%
Hunter%
\ \BBA {} Lange%
}{%
Hunter%
\ \BBA {} Lange%
}{%
{\protect \APACyear {2000}}%
}]{%
MM}
\APACinsertmetastar {%
MM}%
\begin{APACrefauthors}%
Hunter, D\BPBI R.%
\BCBT {}\ \BBA {} Lange, K.%
\end{APACrefauthors}%
\unskip\
\newblock
\APACrefYearMonthDay{2000}{}{}.
\newblock
{\BBOQ}\APACrefatitle {Quantile regression via an {MM} algorithm} {Quantile
  regression via an {MM} algorithm}.{\BBCQ}
\newblock
\APACjournalVolNumPages{Journal of Computational and Graphical
  Statistics}{}{}{60--77}.
\newblock
\begin{APACrefURL} \url{https://doi.org/10.2307/1390613} \end{APACrefURL}
\PrintBackRefs{\CurrentBib}

\bibitem [\protect \citeauthoryear {%
Hunter%
\ \BBA {} Lange%
}{%
Hunter%
\ \BBA {} Lange%
}{%
{\protect \APACyear {2004}}%
}]{%
MM-tutorial}
\APACinsertmetastar {%
MM-tutorial}%
\begin{APACrefauthors}%
Hunter, D\BPBI R.%
\BCBT {}\ \BBA {} Lange, K.%
\end{APACrefauthors}%
\unskip\
\newblock
\APACrefYearMonthDay{2004}{}{}.
\newblock
{\BBOQ}\APACrefatitle {A Tutorial on {MM} Algorithms} {A tutorial on {MM}
  algorithms}.{\BBCQ}
\newblock
\APACjournalVolNumPages{The American Statistician}{58}{}{30--37}.
\newblock
\begin{APACrefURL} \url{https://doi.org/10.1198/0003130042836} \end{APACrefURL}
\PrintBackRefs{\CurrentBib}

\bibitem [\protect \citeauthoryear {%
Jeong%
, N{\'e}da%
\BCBL {}\ \BBA {} Barab{\'a}si%
}{%
Jeong%
\ \protect \BOthers {.}}{%
{\protect \APACyear {2003}}%
}]{%
jeong}
\APACinsertmetastar {%
jeong}%
\begin{APACrefauthors}%
Jeong, H.%
, N{\'e}da, Z.%
\BCBL {}\ \BBA {} Barab{\'a}si, A.%
\end{APACrefauthors}%
\unskip\
\newblock
\APACrefYearMonthDay{2003}{}{}.
\newblock
{\BBOQ}\APACrefatitle {Measuring preferential attachment in evolving networks}
  {Measuring preferential attachment in evolving networks}.{\BBCQ}
\newblock
\APACjournalVolNumPages{Europhysics Letters}{61}{61}{567--572}.
\newblock
\begin{APACrefURL} \url{https://doi.org/10.1209/epl/i2003-00166-9}
  \end{APACrefURL}
\PrintBackRefs{\CurrentBib}

\bibitem [\protect \citeauthoryear {%
Ji%
\ \BBA {} Jin%
}{%
Ji%
\ \BBA {} Jin%
}{%
{\protect \APACyear {2016}}%
}]{%
stats_dataset}
\APACinsertmetastar {%
stats_dataset}%
\begin{APACrefauthors}%
Ji, P.%
\BCBT {}\ \BBA {} Jin, J.%
\end{APACrefauthors}%
\unskip\
\newblock
\APACrefYearMonthDay{2016}{}{}.
\newblock
{\BBOQ}\APACrefatitle {Coauthorship and citation networks for statisticians}
  {Coauthorship and citation networks for statisticians}.{\BBCQ}
\newblock
\APACjournalVolNumPages{The Annals of Applied Statistics}{10}{4}{1779--1812}.
\newblock
\begin{APACrefURL} \url{https://doi.org/10.1214/15-AOAS896} \end{APACrefURL}
\PrintBackRefs{\CurrentBib}

\bibitem [\protect \citeauthoryear {%
Jones%
, Wuchty%
\BCBL {}\ \BBA {} Uzzi%
}{%
Jones%
\ \protect \BOthers {.}}{%
{\protect \APACyear {2008}}%
}]{%
collaboration_science_1}
\APACinsertmetastar {%
collaboration_science_1}%
\begin{APACrefauthors}%
Jones, B\BPBI F.%
, Wuchty, S.%
\BCBL {}\ \BBA {} Uzzi, B.%
\end{APACrefauthors}%
\unskip\
\newblock
\APACrefYearMonthDay{2008}{}{}.
\newblock
{\BBOQ}\APACrefatitle {Multi-University Research Teams: Shifting Impact,
  Geography, and Stratification in Science} {Multi-university research teams:
  Shifting impact, geography, and stratification in science}.{\BBCQ}
\newblock
\APACjournalVolNumPages{Science}{322}{5905}{1259--1262}.
\newblock
\begin{APACrefURL} \url{https://doi.org/10.1126/science.1158357}
  \end{APACrefURL}
\PrintBackRefs{\CurrentBib}

\bibitem [\protect \citeauthoryear {%
Krapivsky%
, Rodgers%
\BCBL {}\ \BBA {} Redner%
}{%
Krapivsky%
\ \protect \BOthers {.}}{%
{\protect \APACyear {2001}}%
}]{%
krapi}
\APACinsertmetastar {%
krapi}%
\begin{APACrefauthors}%
Krapivsky, P.%
, Rodgers, G.%
\BCBL {}\ \BBA {} Redner, S.%
\end{APACrefauthors}%
\unskip\
\newblock
\APACrefYearMonthDay{2001}{}{}.
\newblock
{\BBOQ}\APACrefatitle {Organization of growing random networks} {Organization
  of growing random networks}.{\BBCQ}
\newblock
\APACjournalVolNumPages{Physical Review E}{}{}{066123}.
\newblock
\begin{APACrefURL} \url{https://doi.org/10.1103/PhysRevE.63.066123}
  \end{APACrefURL}
\PrintBackRefs{\CurrentBib}

\bibitem [\protect \citeauthoryear {%
Krivitsky%
\ \BBA {} Handcock%
}{%
Krivitsky%
\ \BBA {} Handcock%
}{%
{\protect \APACyear {2019}}%
}]{%
tergm}
\APACinsertmetastar {%
tergm}%
\begin{APACrefauthors}%
Krivitsky, P\BPBI N.%
\BCBT {}\ \BBA {} Handcock, M\BPBI S.%
\end{APACrefauthors}%
\unskip\
\newblock
\APACrefYearMonthDay{2019}{}{}.
\newblock
{\BBOQ}\APACrefatitle {tergm: Fit, Simulate and Diagnose Models for Network
  Evolution Based on Exponential-Family Random Graph Models} {tergm: Fit,
  simulate and diagnose models for network evolution based on
  exponential-family random graph models}{\BBCQ}\ [\bibcomputersoftwaremanual].
\newblock
\begin{APACrefURL} \url{https://CRAN.R-project.org/package=tergm}
  \end{APACrefURL}
\newblock
\APACrefnote{R package version 3.6.0}
\PrintBackRefs{\CurrentBib}

\bibitem [\protect \citeauthoryear {%
Kronegger%
, Mali%
, Ferligoj%
\BCBL {}\ \BBA {} Doreian%
}{%
Kronegger%
\ \protect \BOthers {.}}{%
{\protect \APACyear {2012}}%
}]{%
rsiena_slovenian_2012}
\APACinsertmetastar {%
rsiena_slovenian_2012}%
\begin{APACrefauthors}%
Kronegger, L.%
, Mali, F.%
, Ferligoj, A.%
\BCBL {}\ \BBA {} Doreian, P.%
\end{APACrefauthors}%
\unskip\
\newblock
\APACrefYearMonthDay{2012}{}{}.
\newblock
{\BBOQ}\APACrefatitle {Collaboration structures in Slovenian scientific
  communities} {Collaboration structures in slovenian scientific
  communities}.{\BBCQ}
\newblock
\APACjournalVolNumPages{Scientometrics}{90}{2}{631--647}.
\newblock
\begin{APACrefURL} \url{https://doi.org/10.1007/s11192-011-0493-8}
  \end{APACrefURL}
\PrintBackRefs{\CurrentBib}

\bibitem [\protect \citeauthoryear {%
Larivi{\`e}re%
, Gingras%
, Sugimoto%
\BCBL {}\ \BBA {} Tsou%
}{%
Larivi{\`e}re%
\ \protect \BOthers {.}}{%
{\protect \APACyear {2015}}%
}]{%
citation_1}
\APACinsertmetastar {%
citation_1}%
\begin{APACrefauthors}%
Larivi{\`e}re, V.%
, Gingras, Y.%
, Sugimoto, C\BPBI R.%
\BCBL {}\ \BBA {} Tsou, A.%
\end{APACrefauthors}%
\unskip\
\newblock
\APACrefYearMonthDay{2015}{}{}.
\newblock
{\BBOQ}\APACrefatitle {Team size matters: Collaboration and scientific impact
  since 1900} {Team size matters: Collaboration and scientific impact since
  1900}.{\BBCQ}
\newblock
\APACjournalVolNumPages{Journal of the Association for Information Science and
  Technology}{66}{7}{1323--1332}.
\newblock
\begin{APACrefURL} \url{https://doi.org/10.1002/asi.23266} \end{APACrefURL}
\PrintBackRefs{\CurrentBib}

\bibitem [\protect \citeauthoryear {%
Liben-Nowell%
\ \BBA {} Kleinberg%
}{%
Liben-Nowell%
\ \BBA {} Kleinberg%
}{%
{\protect \APACyear {2007}}%
}]{%
transitivity_win_PA_in_link_prediction}
\APACinsertmetastar {%
transitivity_win_PA_in_link_prediction}%
\begin{APACrefauthors}%
Liben-Nowell, D.%
\BCBT {}\ \BBA {} Kleinberg, J.%
\end{APACrefauthors}%
\unskip\
\newblock
\APACrefYearMonthDay{2007}{}{}.
\newblock
{\BBOQ}\APACrefatitle {The link-prediction problem for social networks} {The
  link-prediction problem for social networks}.{\BBCQ}
\newblock
\APACjournalVolNumPages{Journal of the American Society for Information Science
  and Technology}{58}{7}{1019--1031}.
\newblock
\begin{APACrefURL} \url{https://doi.org/10.1002/asi.20591} \end{APACrefURL}
\PrintBackRefs{\CurrentBib}

\bibitem [\protect \citeauthoryear {%
J.~Lospinoso%
}{%
J.~Lospinoso%
}{%
{\protect \APACyear {2012}}%
}]{%
stochastic_actor_phd}
\APACinsertmetastar {%
stochastic_actor_phd}%
\begin{APACrefauthors}%
Lospinoso, J.%
\end{APACrefauthors}%
\unskip\
\newblock
\APACrefYear{2012}.
\unskip\
\newblock
\APACrefbtitle {Statistical models for social network dynamics} {Statistical
  models for social network dynamics}\ \APACtypeAddressSchool
  {\BPhD}{UK}{Oxford University}.
\unskip\
\newblock
\begin{APACrefURL} \url{https://ora.ox.ac.uk/objects/ora:6726} \end{APACrefURL}
\PrintBackRefs{\CurrentBib}

\bibitem [\protect \citeauthoryear {%
J\BPBI A.~Lospinoso%
, Schweinberger%
, Snijders%
\BCBL {}\ \BBA {} Ripley%
}{%
J\BPBI A.~Lospinoso%
\ \protect \BOthers {.}}{%
{\protect \APACyear {2011}}%
}]{%
stochastic_actor_model_test_time}
\APACinsertmetastar {%
stochastic_actor_model_test_time}%
\begin{APACrefauthors}%
Lospinoso, J\BPBI A.%
, Schweinberger, M.%
, Snijders, T\BPBI A\BPBI B.%
\BCBL {}\ \BBA {} Ripley, R\BPBI M.%
\end{APACrefauthors}%
\unskip\
\newblock
\APACrefYearMonthDay{2011}{}{}.
\newblock
{\BBOQ}\APACrefatitle {Assessing and accounting for time heterogeneity in
  stochastic actor oriented models} {Assessing and accounting for time
  heterogeneity in stochastic actor oriented models}.{\BBCQ}
\newblock
\APACjournalVolNumPages{Advances in Data Analysis and
  Classification}{5}{2}{147--176}.
\newblock
\begin{APACrefURL} \url{https://doi.org/10.1007/s11634-010-0076-1}
  \end{APACrefURL}
\PrintBackRefs{\CurrentBib}

\bibitem [\protect \citeauthoryear {%
Massen%
\ \BBA {} Jonathan%
}{%
Massen%
\ \BBA {} Jonathan%
}{%
{\protect \APACyear {2007}}%
}]{%
massen}
\APACinsertmetastar {%
massen}%
\begin{APACrefauthors}%
Massen, C.%
\BCBT {}\ \BBA {} Jonathan, P.%
\end{APACrefauthors}%
\unskip\
\newblock
\APACrefYearMonthDay{2007}{}{}.
\newblock
{\BBOQ}\APACrefatitle {Preferential attachment during the evolution of a
  potential energy landscape} {Preferential attachment during the evolution of
  a potential energy landscape}.{\BBCQ}
\newblock
\APACjournalVolNumPages{The Journal of Chemical Physics}{127}{}{114306}.
\newblock
\begin{APACrefURL} \url{https://doi.org/10.1063/1.2773721} \end{APACrefURL}
\PrintBackRefs{\CurrentBib}

\bibitem [\protect \citeauthoryear {%
Medo%
}{%
Medo%
}{%
{\protect \APACyear {2014}}%
}]{%
medo_time_varying}
\APACinsertmetastar {%
medo_time_varying}%
\begin{APACrefauthors}%
Medo, M.%
\end{APACrefauthors}%
\unskip\
\newblock
\APACrefYearMonthDay{2014}{}{}.
\newblock
{\BBOQ}\APACrefatitle {Statistical validation of high-dimensional models of
  growing networks} {Statistical validation of high-dimensional models of
  growing networks}.{\BBCQ}
\newblock
\APACjournalVolNumPages{Physical Review E}{89}{}{032801}.
\newblock
\begin{APACrefURL} \url{https://doi.org/10.1103/PhysRevE.89.032801}
  \end{APACrefURL}
\PrintBackRefs{\CurrentBib}

\bibitem [\protect \citeauthoryear {%
Medo%
, Cimini%
\BCBL {}\ \BBA {} Gualdi%
}{%
Medo%
\ \protect \BOthers {.}}{%
{\protect \APACyear {2011}}%
}]{%
temporal}
\APACinsertmetastar {%
temporal}%
\begin{APACrefauthors}%
Medo, M.%
, Cimini, G.%
\BCBL {}\ \BBA {} Gualdi, S.%
\end{APACrefauthors}%
\unskip\
\newblock
\APACrefYearMonthDay{2011}{}{}.
\newblock
{\BBOQ}\APACrefatitle {Temporal Effects in the Growth of Networks} {Temporal
  effects in the growth of networks}.{\BBCQ}
\newblock
\APACjournalVolNumPages{Physical Review Letter}{107}{}{238701}.
\newblock
\begin{APACrefURL} \url{https://doi.org/10.1103/PhysRevLett.107.238701}
  \end{APACrefURL}
\PrintBackRefs{\CurrentBib}

\bibitem [\protect \citeauthoryear {%
Merton%
}{%
Merton%
}{%
{\protect \APACyear {1968}}%
}]{%
matthew_effect}
\APACinsertmetastar {%
matthew_effect}%
\begin{APACrefauthors}%
Merton, R\BPBI K.%
\end{APACrefauthors}%
\unskip\
\newblock
\APACrefYearMonthDay{1968}{}{}.
\newblock
{\BBOQ}\APACrefatitle {{T}he {M}atthew Effect in Science} {{T}he {M}atthew
  effect in science}.{\BBCQ}
\newblock
\APACjournalVolNumPages{Science}{159}{3810}{56--63}.
\newblock
\begin{APACrefURL} \url{https://doi.org/10.1126/science.159.3810.56}
  \end{APACrefURL}
\PrintBackRefs{\CurrentBib}

\bibitem [\protect \citeauthoryear {%
Milojevi{\'c}%
}{%
Milojevi{\'c}%
}{%
{\protect \APACyear {2010}}%
}]{%
miloj}
\APACinsertmetastar {%
miloj}%
\begin{APACrefauthors}%
Milojevi{\'c}, S.%
\end{APACrefauthors}%
\unskip\
\newblock
\APACrefYearMonthDay{2010}{}{}.
\newblock
{\BBOQ}\APACrefatitle {Modes of collaboration in modern science: Beyond power
  laws and preferential attachment} {Modes of collaboration in modern science:
  Beyond power laws and preferential attachment}.{\BBCQ}
\newblock
\APACjournalVolNumPages{Journal of the American Society for Information Science
  and Technology}{61}{7}{1410--1423}.
\newblock
\begin{APACrefURL} \url{https://doi.org/10.1002/asi.21331} \end{APACrefURL}
\PrintBackRefs{\CurrentBib}

\bibitem [\protect \citeauthoryear {%
Newman%
}{%
Newman%
}{%
{\protect \APACyear {2001}}%
{\protect \APACexlab {{\protect \BCnt {1}}}}}]{%
newman2001clustering}
\APACinsertmetastar {%
newman2001clustering}%
\begin{APACrefauthors}%
Newman, M\BPBI E\BPBI J.%
\end{APACrefauthors}%
\unskip\
\newblock
\APACrefYearMonthDay{2001{\protect \BCnt {1}}}{}{}.
\newblock
{\BBOQ}\APACrefatitle {Clustering and preferential attachment in growing
  networks} {Clustering and preferential attachment in growing
  networks}.{\BBCQ}
\newblock
\APACjournalVolNumPages{Physical Review E}{64}{2}{025102}.
\newblock
\begin{APACrefURL} \url{https://doi.org/10.1103/PhysRevE.64.025102}
  \end{APACrefURL}
\PrintBackRefs{\CurrentBib}

\bibitem [\protect \citeauthoryear {%
Newman%
}{%
Newman%
}{%
{\protect \APACyear {2001}}%
{\protect \APACexlab {{\protect \BCnt {2}}}}}]{%
Newman_coauthornet1}
\APACinsertmetastar {%
Newman_coauthornet1}%
\begin{APACrefauthors}%
Newman, M\BPBI E\BPBI J.%
\end{APACrefauthors}%
\unskip\
\newblock
\APACrefYearMonthDay{2001{\protect \BCnt {2}}}{}{}.
\newblock
{\BBOQ}\APACrefatitle {The structure of scientific collaboration networks} {The
  structure of scientific collaboration networks}.{\BBCQ}
\newblock
\APACjournalVolNumPages{Proceedings of the National Academy of
  Sciences}{98}{2}{404--409}.
\newblock
\begin{APACrefURL} \url{https://doi.org/10.1073/pnas.98.2.404} \end{APACrefURL}
\PrintBackRefs{\CurrentBib}

\bibitem [\protect \citeauthoryear {%
Newman%
}{%
Newman%
}{%
{\protect \APACyear {2005}}%
}]{%
newman_powerlaw}
\APACinsertmetastar {%
newman_powerlaw}%
\begin{APACrefauthors}%
Newman, M\BPBI E\BPBI J.%
\end{APACrefauthors}%
\unskip\
\newblock
\APACrefYearMonthDay{2005}{}{}.
\newblock
{\BBOQ}\APACrefatitle {Power laws, {P}areto distributions and {Z}ipf's law}
  {Power laws, {P}areto distributions and {Z}ipf's law}.{\BBCQ}
\newblock
\APACjournalVolNumPages{Contemporary Physics}{46}{}{323--351}.
\newblock
\begin{APACrefURL} \url{https://doi.org/10.1080/00107510500052444}
  \end{APACrefURL}
\PrintBackRefs{\CurrentBib}

\bibitem [\protect \citeauthoryear {%
Pham%
, Sheridan%
\BCBL {}\ \BBA {} Shimodaira%
}{%
Pham%
\ \protect \BOthers {.}}{%
{\protect \APACyear {2015}}%
}]{%
pham2}
\APACinsertmetastar {%
pham2}%
\begin{APACrefauthors}%
Pham, T.%
, Sheridan, P.%
\BCBL {}\ \BBA {} Shimodaira, H.%
\end{APACrefauthors}%
\unskip\
\newblock
\APACrefYearMonthDay{2015}{}{}.
\newblock
{\BBOQ}\APACrefatitle {{PAFit}: {a} Statistical Method for Measuring
  Preferential Attachment in Temporal Complex Networks} {{PAFit}: {a}
  statistical method for measuring preferential attachment in temporal complex
  networks}.{\BBCQ}
\newblock
\APACjournalVolNumPages{PLOS ONE}{}{9}{e0137796}.
\newblock
\begin{APACrefURL} \url{https://doi.org/10.1371/journal.pone.0137796}
  \end{APACrefURL}
\PrintBackRefs{\CurrentBib}

\bibitem [\protect \citeauthoryear {%
Pham%
, Sheridan%
\BCBL {}\ \BBA {} Shimodaira%
}{%
Pham%
\ \protect \BOthers {.}}{%
{\protect \APACyear {2016}}%
}]{%
pham3}
\APACinsertmetastar {%
pham3}%
\begin{APACrefauthors}%
Pham, T.%
, Sheridan, P.%
\BCBL {}\ \BBA {} Shimodaira, H.%
\end{APACrefauthors}%
\unskip\
\newblock
\APACrefYearMonthDay{2016}{}{}.
\newblock
{\BBOQ}\APACrefatitle {{J}oint Estimation of Preferential Attachment and Node
  Fitness in Growing Complex Networks} {{J}oint estimation of preferential
  attachment and node fitness in growing complex networks}.{\BBCQ}
\newblock
\APACjournalVolNumPages{Scientific Reports}{6}{}{}.
\newblock
\begin{APACrefURL} \url{https://doi.org/10.1038/srep32558} \end{APACrefURL}
\PrintBackRefs{\CurrentBib}

\bibitem [\protect \citeauthoryear {%
Pham%
, Sheridan%
\BCBL {}\ \BBA {} Shimodaira%
}{%
Pham%
\ \protect \BOthers {.}}{%
{\protect \APACyear {to appear}}%
}]{%
pham_jss}
\APACinsertmetastar {%
pham_jss}%
\begin{APACrefauthors}%
Pham, T.%
, Sheridan, P.%
\BCBL {}\ \BBA {} Shimodaira, H.%
\end{APACrefauthors}%
\unskip\
\newblock
\APACrefYearMonthDay{to appear}{}{}.
\newblock
{\BBOQ}\APACrefatitle {{PAFit}: an {R} Package for Estimating Preferential
  Attachment and Node Fitness in Temporal Complex Networks} {{PAFit}: an {R}
  package for estimating preferential attachment and node fitness in temporal
  complex networks}.{\BBCQ}
\newblock
\APACjournalVolNumPages{Journal of Statistical Software}{}{}{}.
\newblock
\begin{APACrefURL} \url{https://arxiv.org/abs/1704.06017} \end{APACrefURL}
\PrintBackRefs{\CurrentBib}

\bibitem [\protect \citeauthoryear {%
Price%
}{%
Price%
}{%
{\protect \APACyear {1965}}%
}]{%
price2}
\APACinsertmetastar {%
price2}%
\begin{APACrefauthors}%
Price, D\BPBI d\BPBI S.%
\end{APACrefauthors}%
\unskip\
\newblock
\APACrefYearMonthDay{1965}{}{}.
\newblock
{\BBOQ}\APACrefatitle {Networks of Scientific Papers} {Networks of scientific
  papers}.{\BBCQ}
\newblock
\APACjournalVolNumPages{Science}{149}{3683}{510--515}.
\newblock
\begin{APACrefURL} \url{https://doi.org/10.1126/science.149.3683.510}
  \end{APACrefURL}
\PrintBackRefs{\CurrentBib}

\bibitem [\protect \citeauthoryear {%
Price%
}{%
Price%
}{%
{\protect \APACyear {1976}}%
}]{%
price}
\APACinsertmetastar {%
price}%
\begin{APACrefauthors}%
Price, D\BPBI d\BPBI S.%
\end{APACrefauthors}%
\unskip\
\newblock
\APACrefYearMonthDay{1976}{}{}.
\newblock
{\BBOQ}\APACrefatitle {A general theory of bibliometric and other cumulative
  advantage processes} {A general theory of bibliometric and other cumulative
  advantage processes}.{\BBCQ}
\newblock
\APACjournalVolNumPages{Journal of the American Society for Information
  Science}{27}{5}{292--306}.
\newblock
\begin{APACrefURL} \url{https://doi.org/10.1002/asi.4630270505}
  \end{APACrefURL}
\PrintBackRefs{\CurrentBib}

\bibitem [\protect \citeauthoryear {%
Ripley%
, Snijders%
, Boda%
, V\"or\"os%
\BCBL {}\ \BBA {} Preciado%
}{%
Ripley%
\ \protect \BOthers {.}}{%
{\protect \APACyear {2018}}%
}]{%
rsiena}
\APACinsertmetastar {%
rsiena}%
\begin{APACrefauthors}%
Ripley, R\BPBI M.%
, Snijders, T\BPBI A.%
, Boda, Z.%
, V\"or\"os, A.%
\BCBL {}\ \BBA {} Preciado, P.%
\end{APACrefauthors}%
\unskip\
\newblock
\APACrefYearMonthDay{2018}{}{}.
\newblock
{\BBOQ}\APACrefatitle {Manual for SIENA version 4.0 (version May 24, 2018)}
  {Manual for siena version 4.0 (version may 24, 2018)}{\BBCQ}\
  [\bibcomputersoftwaremanual].
\newblock
\begin{APACrefURL} \url{http://www.stats.ox.ac.uk/~snijders/siena/}
  \end{APACrefURL}
\PrintBackRefs{\CurrentBib}

\bibitem [\protect \citeauthoryear {%
Ronda-Pupo%
\ \BBA {} Pham%
}{%
Ronda-Pupo%
\ \BBA {} Pham%
}{%
{\protect \APACyear {2018}}%
}]{%
pham4}
\APACinsertmetastar {%
pham4}%
\begin{APACrefauthors}%
Ronda-Pupo, G\BPBI A.%
\BCBT {}\ \BBA {} Pham, T.%
\end{APACrefauthors}%
\unskip\
\newblock
\APACrefYearMonthDay{2018}{}{}.
\newblock
{\BBOQ}\APACrefatitle {The evolutions of the rich get richer and the fit get
  richer phenomena in scholarly networks: the case of the strategic management
  journal} {The evolutions of the rich get richer and the fit get richer
  phenomena in scholarly networks: the case of the strategic management
  journal}.{\BBCQ}
\newblock
\APACjournalVolNumPages{Scientometrics}{}{}{}.
\newblock
\begin{APACrefURL} \url{https://doi.org/10.1007/s11192-018-2761-3}
  \end{APACrefURL}
\PrintBackRefs{\CurrentBib}

\bibitem [\protect \citeauthoryear {%
Simon%
}{%
Simon%
}{%
{\protect \APACyear {1955}}%
}]{%
simon}
\APACinsertmetastar {%
simon}%
\begin{APACrefauthors}%
Simon, H\BPBI A.%
\end{APACrefauthors}%
\unskip\
\newblock
\APACrefYearMonthDay{1955}{}{}.
\newblock
{\BBOQ}\APACrefatitle {On a Class of Skew Distribution Functions} {On a class
  of skew distribution functions}.{\BBCQ}
\newblock
\APACjournalVolNumPages{Biometrika}{42}{3/4}{425--440}.
\newblock
\begin{APACrefURL} \url{https://doi.org/10.1093/biomet/42.3-4.425}
  \end{APACrefURL}
\PrintBackRefs{\CurrentBib}

\bibitem [\protect \citeauthoryear {%
Snijders%
}{%
Snijders%
}{%
{\protect \APACyear {2001}}%
}]{%
stochastic_actor_1}
\APACinsertmetastar {%
stochastic_actor_1}%
\begin{APACrefauthors}%
Snijders, T\BPBI A.%
\end{APACrefauthors}%
\unskip\
\newblock
\APACrefYearMonthDay{2001}{}{}.
\newblock
{\BBOQ}\APACrefatitle {The Statistical Evaluation of Social Network Dynamics}
  {The statistical evaluation of social network dynamics}.{\BBCQ}
\newblock
\APACjournalVolNumPages{Sociological Methodology}{31}{1}{361--395}.
\newblock
\begin{APACrefURL} \url{https://doi.org/10.1111/0081-1750.00099}
  \end{APACrefURL}
\PrintBackRefs{\CurrentBib}

\bibitem [\protect \citeauthoryear {%
Snijders%
}{%
Snijders%
}{%
{\protect \APACyear {2017}}%
}]{%
stochastic_actor_2}
\APACinsertmetastar {%
stochastic_actor_2}%
\begin{APACrefauthors}%
Snijders, T\BPBI A.%
\end{APACrefauthors}%
\unskip\
\newblock
\APACrefYearMonthDay{2017}{}{}.
\newblock
{\BBOQ}\APACrefatitle {Stochastic Actor-Oriented Models for Network Dynamics}
  {Stochastic actor-oriented models for network dynamics}.{\BBCQ}
\newblock
\APACjournalVolNumPages{Annual Review of Statistics and Its
  Application}{4}{1}{343--363}.
\newblock
\begin{APACrefURL}
  \url{https://doi.org/10.1146/annurev-statistics-060116-054035}
  \end{APACrefURL}
\PrintBackRefs{\CurrentBib}

\bibitem [\protect \citeauthoryear {%
Wang%
, Yu%
\BCBL {}\ \BBA {} Yu%
}{%
Wang%
\ \protect \BOthers {.}}{%
{\protect \APACyear {2008}}%
}]{%
wang_measuring}
\APACinsertmetastar {%
wang_measuring}%
\begin{APACrefauthors}%
Wang, M.%
, Yu, G.%
\BCBL {}\ \BBA {} Yu, D.%
\end{APACrefauthors}%
\unskip\
\newblock
\APACrefYearMonthDay{2008}{}{}.
\newblock
{\BBOQ}\APACrefatitle {Measuring the preferential attachment mechanism in
  citation networks} {Measuring the preferential attachment mechanism in
  citation networks}.{\BBCQ}
\newblock
\APACjournalVolNumPages{Physica A: Statistical Mechanics and its
  Applications}{387}{18}{4692--4698}.
\newblock
\begin{APACrefURL} \url{https://doi.org/10.1016/j.physa.2008.03.017}
  \end{APACrefURL}
\PrintBackRefs{\CurrentBib}

\bibitem [\protect \citeauthoryear {%
Yule%
}{%
Yule%
}{%
{\protect \APACyear {1925}}%
}]{%
yule}
\APACinsertmetastar {%
yule}%
\begin{APACrefauthors}%
Yule, G\BPBI U.%
\end{APACrefauthors}%
\unskip\
\newblock
\APACrefYearMonthDay{1925}{}{}.
\newblock
{\BBOQ}\APACrefatitle {A Mathematical Theory of Evolution, Based on the
  Conclusions of {D}r. {J}.{C}. {W}illis,{F}.{R}.{S}.} {A mathematical theory
  of evolution, based on the conclusions of {D}r. {J}.{C}.
  {W}illis,{F}.{R}.{S}.}{\BBCQ}
\newblock
\APACjournalVolNumPages{Philosophical Transactions of the Royal Society of
  London B: Biological Sciences}{213}{402--410}{21--87}.
\newblock
\begin{APACrefURL} \url{https://doi.org/10.1098/rstb.1925.0002}
  \end{APACrefURL}
\PrintBackRefs{\CurrentBib}

\bibitem [\protect \citeauthoryear {%
Zeng%
\ \protect \BOthers {.}}{%
Zeng%
\ \protect \BOthers {.}}{%
{\protect \APACyear {2017}}%
}]{%
scisci2}
\APACinsertmetastar {%
scisci2}%
\begin{APACrefauthors}%
Zeng, A.%
, Shen, Z.%
, Zhou, J.%
, Wu, J.%
, Fan, Y.%
, Wang, Y.%
\BCBL {}\ \BBA {} Stanley, H\BPBI E.%
\end{APACrefauthors}%
\unskip\
\newblock
\APACrefYearMonthDay{2017}{}{}.
\newblock
{\BBOQ}\APACrefatitle {The science of science: From the perspective of complex
  systems} {The science of science: From the perspective of complex
  systems}.{\BBCQ}
\newblock
\APACjournalVolNumPages{Physics Reports}{714--715}{}{1--73}.
\newblock
\begin{APACrefURL} \url{https://doi.org/10.1016/j.physrep.2017.10.001}
  \end{APACrefURL}
\PrintBackRefs{\CurrentBib}

\bibitem [\protect \citeauthoryear {%
Zinilli%
}{%
Zinilli%
}{%
{\protect \APACyear {2016}}%
}]{%
rsiena_transitivity_2}
\APACinsertmetastar {%
rsiena_transitivity_2}%
\begin{APACrefauthors}%
Zinilli, A.%
\end{APACrefauthors}%
\unskip\
\newblock
\APACrefYearMonthDay{2016}{}{}.
\newblock
{\BBOQ}\APACrefatitle {Competitive project funding and dynamic complex
  networks: evidence from {P}rojects of {N}ational {I}nterest ({PRIN})}
  {Competitive project funding and dynamic complex networks: evidence from
  {P}rojects of {N}ational {I}nterest ({PRIN})}.{\BBCQ}
\newblock
\APACjournalVolNumPages{Scientometrics}{108}{2}{633--652}.
\newblock
\begin{APACrefURL} \url{https://doi.org/10.1007/s11192-016-1976-4}
  \end{APACrefURL}
\PrintBackRefs{\CurrentBib}

\end{thebibliography}

\appendix
\renewcommand{\theequation}{\thesection.\arabic{equation}}
\setcounter{equation}{0}
\section{An MM algorithm for estimating the non-parametric PA and transitivity functions}\label{sec:appendix_1}
To maximize the partial log-likelihood function $l(\bm{A}, \bm{B})$ in Eq.~(\ref{eq:likelihood2}), we derive an instance of the Minorize-Maximization algorithms~\cite{MM}. Denote $A_{k}^{(q)}$ the value of $A_k$ at iteration $q$ ($q \ge 0$), and $\bm{A}^{(q)} = [A_0^{(q)}, A_1^{(q)},\ldots, A_{k_{max}}^{(q)}]$ the value of $\bm{A}$ at that iteration. Define $B_{b}^{(q)}$ and $\bm{B}^{(q)}$ in a similar way. Starting from some initial values $(\bm{A}^{0},\bm{B}^{0})$ at iteration $q = 0$, we want to compute $(\bm{A}^{(q+1)},\bm{B}^{(q+1)})$ from $(\bm{A}^{(q)},\bm{B}^{(q)})$. In MM algorithms, one derives such update formulas by first finding a surrogate function $Q(\bm{A}, \bm{B})$ that satisfies $l(\bm{A}, \bm{B}) \geq Q(\bm{A}, \bm{B}), \forall \bm{A},\bm{B}$ and $l(\bm{A}^{(q)}, \bm{B}^{(q)}) = Q(\bm{A}^{(q)}, \bm{B}^{(q)})$, and then maximize the surrogate function. One can prove that, if $(\bm{A}^{q+1},\bm{B}^{q+1})$ maximizes $Q(\bm{A}, \bm{B})$, then $l(\bm{A}^{(q + 1)}, \bm{B}^{(q + 1)}) \geq l(\bm{A}^{(q)}, \bm{B}^{(q)})$, i.e., the objective function is increased monotonically per iteration. Since there can be many surrogate functions that satisfy the conditions, the main indicator for evaluating a particular $Q(\bm{A}, \bm{B})$ is how easily we can maximize it.

Based on previous works~\cite{pham2,pham3}, the following function is a surrogate function of $l$:
\begin{align}
Q'(\bm{A}, \bm{B})=& \sum_{t=0}^{T}\sum_{i=0 }^{K}\sum_{j=i}^{K}\sum_{l=0}^{B} m_{i,j,l}(t) \log{A_{i}A_{j}B_{l}} - \sum_{t=0}^{T}m(t) \log{ \left( \sum_{i=0 }^{K}\sum_{j=i}^{K}\sum_{l=0}^{B} n_{i,j,l}(t) {A_{i}^{(q)}A_{j}^{(q)}B_{l}^{(q)}} \right) }\nonumber\\
- & \sum_{t=0}^{T}m(t) \frac{  \sum_{i =0}^{K}\sum_{j=i}^{K}\sum_{l=0}^{B} n_{i,j,l}(t) {A_{i}A_{j}B_{l}} }{\sum_{i=0 }^{K}\sum_{j=i}^{K}\sum_{l=0}^{B} n_{i,j,l}(t) {A_{i}^{(q)}A_{j}^{(q)}B_{l}^{(q)}}  } \  + \   \sum_{t=0}^{T}m(t), \label{eq:supp_first_surrogate}
\end{align}
where $K := k_{max}$ and $B := b_{max}$.
\\
The product $A_iA_jB_l$ in the numerator of the third term in the r.h.s. of Eq.~(\ref{eq:supp_first_surrogate}) prevents parallel updating of $\bm{A}$ and $\bm{B}$. One way to deal with this product is to apply the AM-GM inequality~\cite{MM-tutorial}:
\begin{align}
-A_iA_jB_l \geq & -\frac{1}{2} \left(  \frac{A_j^{(q)}}{A_i^{(q)}} A_i^2 + \frac{A_i^{(q)}}{A_j^{(q)}} A_j^2  \right) B_l \nonumber\\
\geq & -\frac{1}{4} \left(  \frac{A_j^{(q)}B_l^{(q)}}{\left( A_i^{(q)}  \right)^3 } A_i^4 + \frac{A_i^{(q)}B_l^{(q)}}{\left( A_j^{(q)}  \right)^3 } A_j^4 \right) - \frac{1}{2}\frac{A_i^{(q)}A_j^{(q)}}{B_l^{(q)}} B_l^2.\nonumber
\end{align}
Plugging this inequality to Eq.~(\ref{eq:supp_first_surrogate}), we obtain the final surrogate function:
\begin{align}
Q(\bm{A}, \bm{B}) = &\sum_{t=0}^{T}\sum_{i=0 }^{K}\sum_{j=i}^{K}\sum_{l=0}^{B} m_{i,j,l}(t) \log{A_{i}A_{j}B_{l}} - \sum_{t=0}^{T}m(t) \log{ \left( \sum_{i=0 }^{K}\sum_{j=i}^{K}\sum_{l=0}^{B} n_{i,j,l}(t) {A_{i}^{(q)}A_{j}^{(q)}B_{l}^{(q)}} \right) }\nonumber\\
- & \sum_{t=0}^{T}m(t) \frac{  \sum_{i =0}^{K}\sum_{j=i}^{K}\sum_{l=0}^{B} n_{i,j,l}(t) {\left( \frac{1}{4} \left(  \frac{A_j^{(q)}B_l^{(q)}}{\left( A_i^{(q)}  \right)^3 } A_i^4 + \frac{A_i^{(q)}B_l^{(q)}}{\left( A_j^{(q)}  \right)^3 } A_j^4 \right) + \frac{1}{2}\frac{A_i^{(q)}A_j^{(q)}}{B_l^{(q)}} B_l^2 \right)} }{\sum_{i=0 }^{K}\sum_{j=i}^{K}\sum_{l=0}^{B} n_{i,j,l}(t) {A_{i}^{(q)}A_{j}^{(q)}B_{l}^{(q)}}  } \  + \   \sum_{t=0}^{T}m(t). \nonumber
\end{align}
Solving $\frac{\partial Q(\bm{A}, \bm{B})}{\partial A_k} = 0$ and $\frac{\partial Q(\bm{A}, \bm{B})}{\partial B_b} = 0$, we obtain the following closed-form formulas:
\small
\begin{align}
A_k^{(q+1)} &= \sqrt[4]{\frac{\sum_{t=0}^{T}  \sum_{i = 0 }^{k}{m_{i,k,\cdot}(t)} + \sum_{t=0}^{T}  \sum_{j = k}^{K}{m_{k,j,\cdot}(t)}}{\sum_{t=0}^{T}m(t) \frac{ \sum_{  j = k+1 }^{K}\sum_{l=0}^{B}n_{k,j,l}(t)\frac{A_j^{(q)}B_l^{(q)}}{\left( A_k^{(q)}  \right)^3 } \ + \ \sum_{i = 0 }^{k-1}\sum_{l}^{B}n_{i,k,l}(t) \frac{A_i^{(q)}B_l^{(q)}}{\left( A_k^{(q)}  \right)^3 }  \  + \  \sum_{l=0}^{B}n_{k,k,l}(t)\left(  \frac{A_k^{(q)}B_l^{(q)}}{ \left( A_k^{(q)}  \right)^3 }  + \frac{A_k^{(q)}B_l^{(q)}}{ \left( A_k^{(q)}  \right)^3 } \right) }{\sum_{i=0}^{K}\sum_{j=i}^{K}\sum_{l=0}^{B} n_{i,j,l}(t) {A_{i}^{(q)}A_{j}^{(q)}B_{l}^{(q)}}}}}, \nonumber \\
\nonumber\\
B_b^{(q+1)} &= \sqrt{ \frac{\sum_{t=0}^{T}  m_{\cdot,\cdot,b}(t)}{ \sum_{t=0}^{T}m(t) \frac{ \sum_{i =0 }^{K}\sum_{j =i}^{K}n_{i,j,b}(t)\frac{A_i^{(q)}A_j^{(q)}}{B_b^{(q)}}  }{\sum_{i =0}^{K}\sum_{j=i}^{K}\sum_{l=0}^{B} n_{i,j,l}(t) {A_{i}^{(q)}A_{j}^{(q)}B_{l}^{(q)}}} }}, \nonumber
\end{align}
\small
\normalsize
where $m_{i,k,\cdot}(t):=\sum_{l=0}^{B}m_{i,k,l}(t)$ and $m_{\cdot,\cdot,b}:=\sum_{i = 0}^{K}\sum_{j=i}^{K}m_{i,j,b}(t)$.

Based on these formulas, at each iteration $\bm{A}^{(q+1)}$ and $\bm{B}^{(q+1)}$ can be computed in parallel without solving any additional optimization problems. This enables the method to work with large data-sets. The objective function value $l(\bm{A}^{(q+1)}, \bm{B}^{(q+1)})$, as explained earlier, is guaranteed to be increasing in $q$.
\section{Estimation of the standard deviations of $\hat{h}_{\text{trans}}(t)$ and $\hat{h}_{\text{PA}}(t)$}\label{sec:appendix_2}
We have the following closed-form formula for the variance of the sample variance $h_{\text{trans}}(t)^2$:
$$
\mathbb{V} h_{\text{trans}}(t)^2 = \frac{1}{m(t)}\E (\log_2 B_{b_{ij}} - \E \log_2 B_{b_{ij}})^4 - \frac{(m(t) - 3)s_{\text{trans}}(t)^4}{m(t)(m(t) - 1)}.
$$
The delta method then gives:
$$
sd(h_{\text{trans}}(t)) \approx \frac{1}{2}\left(\E h_{\text{trans}}(t)^2\right)^{-1/2}\sqrt{\mathbb{V} h_{\text{trans}}(t)^2} \approx \frac{1}{2}\left(s_{\text{trans}}(t)\right)^{-1}\sqrt{\mathbb{V} h_{\text{trans}}(t)^2}.
$$
The standard deviation of $\hat{h}_{\text{trans}}(t)$ then can be calculated by plugging $\hat{\bm{A}}$ and $\hat{\bm{B}}$ into the above formula. The standard deviation of $\hat{h}_{\text{PA}}(t)$ follows the same derivation.
\end{document}